\documentclass[preprint,prc,aps,superscriptaddress,
nofootinbib,showpacs]{revtex4}
\usepackage[dvips]{graphicx}
\usepackage{amsfonts, color, float, bbm, amsmath}
\usepackage{amssymb}
\usepackage{bm}

\def\lp {\left( }
\def\rp {\right) }
\def\lb {\left[ }
\def\rb {\right] }
\def\lc {\left\{ }
\def\rc {\right\} }
\def\ra {\rangle }
\def\la {\langle }
\def\ni {\noindent}
\def\nn {\nonumber}

\def\rar {\rightarrow}
\def\lrar {\leftrightarrow}

\def\beq{\begin{equation}}
\def\eeq{\end{equation}}
\def\bea{\begin{eqnarray}}
\def\eea{\end{eqnarray}}

\def\cO{{\cal{O}}}
\def\cT {{\cal{T}}}

\def\a{\alpha}

\def\d{\delta}

\def\e{\epsilon}
\def\g{\gamma}

\def\m{\mu}
\def\n{\nu}

\def\p {\pi}
\def\r{\rho}
\def\s{\sigma}

\def\t {\tau}

\def\bb {\bar{b}}
\def\db {\bar{d}}
\def\pb {\bar p}

\def\ub {\bar u}

\def\sp {\!+\!}
\def\sm {\!-\!}
\def\st {\!\times \!}
\def\cd {\!\cdot\!}

\def\su { \bsig^{(1)} }
\def\sd { \bsig^{(2)} }
\def\str { \bsig^{(3)} }

\def\rtu {x_{31} }
\def\hrtu {\hat{\bx}_{31} }
\def\rdt {x_{23} }
\def\hrdt {\hat{\bx}_{23} }

\def\bnblr {\bnb^{^{\!\!\!\!\!\!\!\!^\leftrightarrow}}}
\def\bnbr {\bnb^{^{\!\!\!\!\!\!\!\!^\rightarrow}}}
\def\bnbl {\bnb^{^{\!\!\!\!\!\!\!\!^\leftarrow}}}

\def\bk {\mbox{\boldmath $k$}}

\def\br {\mbox{\boldmath $r$}}
\def\bR {\mbox{\boldmath $R$}}
\def\bro {\mbox{\boldmath $\rho$}}

\def\bsig {\mbox{\boldmath $\sigma$}}
\def\btau {\mbox{\boldmath $\tau$}}

\def\bnb {\mbox{\boldmath $\nabla$}}

\def\bp {\mbox{\boldmath $p$}}
\def\bP {\mbox{\boldmath $P$}}

\def\bq {\mbox{\boldmath $q$}}
\def\bQ {\mbox{\boldmath $Q$}}
\def\bx {\mbox{\boldmath $x$}}
\def\bX {\mbox{\boldmath $X$}}

\def\bz {\mbox{\boldmath $z$}}

\begin{document}

\title{Two-pion exchange three-nucleon potential:\\ 
$\cO(q^4)$ chiral expansion }

%%%%%%%%%%%%%%%%%%%%%%%%%%%%%%%%%%%%%%%%%%%%%%%%%%%%%%%%%%%%%%%%%%%%%%%%%%%%%
\author{S. Ishikawa}
\affiliation{Department of Physics, Science Research Center, Hosei University, 
2-17-1 Fujimi, Chiyoda, Tokyo 102-8160, Japan
}

\author{M. R. Robilotta }
\affiliation{Instituto de F\'{\i}sica, Universidade de S\~{a}o Paulo, %\\
C.P. 66318, 05315-970, S\~ao Paulo, SP, Brazil}

%%%%%%%%%%%%%%%%%%%%%%%%%%%%%%%%%%%%%%%%%%%%%%%%%%%%%%%%%%%%%%%%%%%%%%%%%%%%%%
\date{\today}

\begin{abstract}

We present the expansion of the two-pion exchange three-nucleon potential
(TPE-3NP) to chiral order $q^4$, which corresponds to a subset of all 
possibilities at this order and is based on the $\pi$N amplitude at $\cO(q^3)$.
Results encompass both numerical corrections to strength coefficients of
previous $\cO(q^3)$ terms and new structures in the profile functions.
The former are typically smaller than 10\% whereas the latter arise from either 
loop functions or non-local gradients acting on the wave function.
The influence of the new TPE-3NP over static and scattering three-body
observables has been assessed and found to be small, as expected
from perturbative corrections.

\end{abstract}


\maketitle

%^^^^^^^^^^^^^^^^^^^^^^^^^^^^^^^^^^^^^^^^^^^^^^^^^^^^^^^^^^^^^^^^^^^
%1111111111111111111111111111111111111111111111111111111111111111111
\section{INTRODUCTION}
\label{sec:intro}

The research programme for nuclear forces, outlined more than fifty 
years ago by Taketani, Nakamura, and Sasaki \cite{TNS}, treats pions and
nucleons as basic degrees of freedom.
This insight proved to be very fruitful. 
On the one hand, it implies the interconnection of all nuclear processes, 
both among themselves and with a class of free reactions.
On the other, it determines a close relationship
between the number of pions involved in a given interaction and its range.
As a consequence, the outer components of nuclear forces are dominated by
just a few basic subamplitudes, describing either single ($N \rar \p N$) or 
multipion ($\p \p \rar \p \p$, $\p N \rar \p N$, $\p N \rar \p \p N$, ...) 
interactions.

Nevertheless, it took a long time for a theoretical tool to be available  
which allows the precise treatment of these amplitudes. 
Nowadays, owing to the development of chiral perturbation theory (ChPT) in  
association with effective lagrangians \cite{W,W3}, the roles of pions 
and nucleons in nuclear forces can be described consistently.
The rationale for this approach is that the quarks $u$ and $d$, which have small masses, dominate low-energy interactions.
One then works with a two-flavor version of QCD and treats their masses 
as perturbations in a chiral symmetric lagrangian.
The systematic inclusion of quark mass contributions is performed by means of 
chiral perturbation theory, which incorporates 
low-energy features of QCD into the nuclear force problem.
In performing perturbative expansions, one uses a typical scale $q$, 
set by either pion four-momenta or nucleon three-momenta, such that $q\ll 1$ GeV.

Nuclear forces are dominated by two-body $(NN)$ interactions and leading
contributions are due to the one-pion exchange potential (OPEP), 
which begins \cite{Bira} at $\cO(q^0)$.
The two-pion exchange potential (TPEP) begins at $\cO(q^2)$ and, at present, 
there are two independent expansions up to $\cO(q^4)$ in the literature, 
based on either heavy baryon \cite{HB} or covariant \cite{HR,HRR} ChPT. 
The TPEP is closely related with the off-shell $\p N$ amplitude and, at this
order, two-loop diagrams involving intermediate $\p \p$ scattering
already begin to contribute. 

In proper three-nucleon ($3N$) interactions, the leading term is due 
to the process known as TPE-3NP, in which the pion emitted 
by a nucleon is scattered before being absorbed by another one.
It has been available since long \cite{TM79,TM81,CDR}, involves only tree-level 
interactions and has the longest possible range.
This contribution begins at $\cO(q^3)$ and consistency with available 
$NN$ forces demands the extension of the chiral series for the 3NP up 
to $\cO(q^4)$.
However, the implementation of this programme is not straightforward, 
since it requires the evaluation of a rather large number of diagrams.
With the purpose of exploring the magnitude of $\cO(q^4)$ effects, 
in this work we concentrate on the particular subset of processes  
which still belong to the TPE-3NP class. 
Our presentation is divided as follows.
In section \ref{sec:formulation} we display the general relationship between the TPE-3NP 
and the $\p N$ amplitude, in order to  discuss how it affects chiral power 
counting in the former.
The $\p N$ amplitude relevant for the $\cO(q^4)$ potential is derived in
section \ref{sec:piN-amplitude} and used to construct the three-body interaction in section \ref{sec:TPE-P}.
We concentrate on numerical changes induced into both potential parameters 
and observables in sections \ref{sec:coefficients} and \ref{sec:results}, whereas conclusions
are presented in section \ref{eq:conclusions}. 
There are also four appendices, dealing with kinematics, $\p N$ subthreshold 
coefficients, loop integrals and non-local terms.
 
%^^^^^^^^^^^^^^^^^^^^^^^^^^^^^^^^^^^^^^^^^^^^^^^^^^^^^^^^^^^^^^^^^^^^^^
%2222222222222222222222222222222222222222222222222222222222222222222
\section{general formulation}
\label{sec:formulation}

Potentials to be used into non-relativistic equations can be derived from 
field theory by means of the $T$-matrix.
In the case of three-nucleon potentials, one starts from the 
non-relativistic transition matrix describing the process  
$N(p_1)\;N(p_2)\;N(p_3) \rar N(p'_1)\;N(p'_2)\;N(p'_3)$, which includes both 
kernels and their iterations.
The former correspond to proper interactions, represented by diagrams which 
cannot be split into two pieces by cutting positive-energy nucleon lines only,
whereas the latter are automatically generated by the dynamical equation.
Therefore, just the kernels, denoted collectively by $\bar{t}_3$, 
are included into the potential.

\begin{figure}[ht]
\begin{center}
%\vspace{-5mm}
%\hspace*{-25mm}
\includegraphics[width=1.0\columnwidth,angle=0]{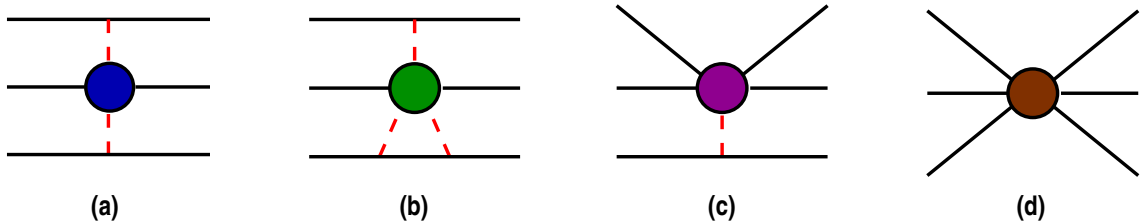}
%\vspace{-15mm}
\caption{(Color online) Classes of three-nucleon forces, where
full and dashed lines represent nucleons and pions respectively;
diagram $(a)$ corresponds to the TPE-3NP.} 
\label{F1}
\end{center}      
\end{figure}

The transformation of a $T$-matrix into a potential depends
on both the dynamical equation adopted and conventions associated 
with off-shell effects.
The latter were discussed in a comprehensive paper by Friar \cite{Fr}.
Here we use the kinematical variables defined in Appendix \ref{secA:kinematics} and relate 
$\bar{t}_3$ to the momentum space potential operator $\hat{W}$ by writing \cite{Y}
\beq
\la \bp'_1, \bp'_2, \bp'_3 \,| \hat{W} |\, \bp_1, \bp_2, \bp_3 \ra = 
- (2\p)^3 \, \d^3(\bP' \sm \bP) \; 
\bar{t}_3 (\bp'_1, \bp'_2, \bp'_3, \bp_1, \bp_2, \bp_3)  \;.
\label{2.1}
\eeq
In configuration space, internal dynamics is described by the function
\bea
&& W(\br', \bro'; \br, \bro) = -\, \lb 2/\sqrt{3} \rb^6
\int \frac{d\bQ_r}{(2\p)^3} \; \frac{d\bQ_\r}{(2\p)^3} 
\; \frac{d\bq_r}{(2\p)^3} \; \frac{d\bq_\r}{(2\p)^3} \;
\nn\\[2mm]
&& \;\;\;\;\; \;\;\;\;\; \times
e^{i \lb \bQ_r \cdot (\br' \sm \br) + \; \bQ_\r \cdot (\bro' \sm \bro)  
+ \bq_r \cdot (\br' \sp \br) /2 + \bq_\r \cdot (\bro' \sp \bro) /2   \rb}
\;\; \bar{t}_3 (\bQ_r, \bQ_\r, \bq_r, \bq_\r)  \;,
\label{2.2}
\eea
\ni
which is to be used in a non-local version of the Schr\"odinger equation:  
\beq
\lb -\, \frac{1}{m}\,\bnb_{r'}^2 - \frac{1}{m}\,\bnb_{\r'}^2 
- \e \, \rb \psi(\br', \bro')
= - \lb \sqrt{3}/2 \rb^3 \int d\br \, d\bro 
\; W(\br', \bro'; \br, \bro) \; \psi (\br, \bro) \;.
\label{2.3}
\eeq
Non-local effects are associated with the variables $\bQ_r$ and $\bQ_\r$.
When these effects are not too strong, they can be represented 
by gradients acting on the wave function and the potential $W$ 
is rewritten as 
\beq
W(\br', \bro'; \br, \bro) = \d^3(\br' \sm \br) \, 
\d^3(\bro' \sm \bro) \lb 2/\sqrt{3} \rb^3 \, V(\br, \bro) \;.
\label{2.4}
\eeq
The two-pion exchange three-nucleon potential is represented in Fig. \ref{F1}a.
It is closely related with the $\p N$ scattering amplitude, 
which is $\cO(q)$ for free pions and becomes $\cO(q^2)$ within the 
three-nucleon system.
As a consequence, the TPE-3NP begins at $\cO(q^3)$ and, at this order,
it also receives contributions from interactions $(c)$ and $(d)$, which have shorter range.
The extension of the chiral series to $\cO(q^4)$ requires both the inclusion 
of single loop effects into processes that already contribute at $\cO(q^3)$
and the evaluation of many new amplitudes, especially those associated with
diagram (b).

%\newpage
%fig.1^^^^^^^^^^^^^^^^^^^^^^^^^^^^^^^^^^^^^^^^^^^^^^^^^^^^^^^^^^^^^^^^^^^^^^^^^
%^^^^^^^^^^^^^^^^^^^^^^^^^^^^^^^^^^^^^^^^^^^^^^^^^^^^^^^^^^^^^^^^^^^^^^^^^^^^^^

%\vspace{-5mm}

\begin{figure}[ht]
\begin{center}
%vspace{-5mm}
%\hspace*{-25mm}
\includegraphics[width=0.40\columnwidth,angle=0]{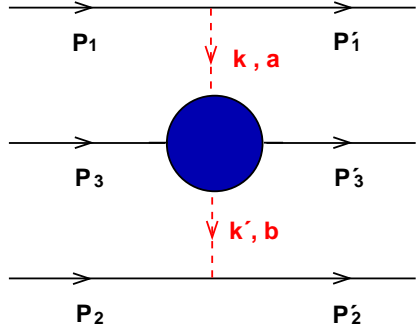}
%\vspace{-4mm}
\caption{(Color online) Two-pion exchange three-nucleon potential.} 
\label{F2}
\end{center}      
\end{figure}

In this paper we concentrate on the particular set of processes which 
belong to the TPE-3NP class, represented by the $T$-matrix $\cT_{\p\p}$ 
and evaluated using the kinematical conditions given in Fig. \ref{F2}.
The coupling of a pion to nucleon $i=(1,2)$ is derived from the usual 
lowest order pseudo-vector lagrangian ${\cal L}^{(1)}$ and 
the Dirac equation yields the equivalent forms for the vertex
\beq
(g_A/2 f_\p) \lb \t \; \ub\, (p' \sm p) \, \g_5 \, u \rb^{(i)} 
= (m g_A/ f_\p) \lb \t \; \ub\, \g_5 \, u \rb^{(i)} \;,
\label{2.5}
\eeq
\ni 
where $g_A$, $f_\p$ and $m$ represent, respectively, 
the axial nucleon decay, 
the pion decay and the nucleon mass.

%\newpage
%fig.2^^^^^^^^^^^^^^^^^^^^^^^^^^^^^^^^^^^^^^^^^^^^^^^^^^^^^^^^^^^^^^^^^^^^^^^^^
%^^^^^^^^^^^^^^^^^^^^^^^^^^^^^^^^^^^^^^^^^^^^^^^^^^^^^^^^^^^^^^^^^^^^^^^^^^^^^^

%\vspace{-5mm}

The amplitude for the intermediate process $\pi^a(k) N(p)\rar\pi^b(k') N(p')$ 
has the isospin structure
\beq
T_{ba} = \d_{ab} \; T^+ +  i\e_{bac}\t_c \; T^-
\label{2.6}
\eeq
\ni
and Fig. \ref{F2} yields
\bea
{\cT}_{\p\p}\!\! & = &\!\!  - \lb \frac{m g_A}{f_\p} \rb^2  
\lb \ub\, \g_5 \, u \rb^{(1)}  \lb \ub\, \g_5 \, u \rb^{(2)}
\frac{1}{k^2 \sm \m^2} \, \frac{1}{k^{'2} \sm \m^2} \,
\nn\\[2mm]
&& \!\! \lb \btau^{(1)} \cd \btau^{(2)}\, T^+ 
- i\, \btau^{(1)}\st \btau^{(2)} \cd \btau^{(3)} \, T^- \rb^{(3)} \;,
\label{2.7}
\eea
\ni 
$\m$ being the pion mass.
Results in Appendix \ref{secA:kinematics} show that $\lb \ub\, \g_5 \, u \rb^{(i)}\rar \cO(q)$, 
whereas pion propagators are $\cO(q^{-2})$. 
As a consequence, in the $\cO(q^4)$ expansion of the potential one needs
${\cT}_{\p\p}$ to $\cO(q)$ and $T^\pm$ to $\cO(q^3)$.
For on-shell nucleons, the sub amplitudes $T^\pm$ can be written as
\beq
T^\pm = \ub(\bp') \lb D^\pm - \;\frac{i}{2m} \s_{\m\n} (p' \sm p)^\m K^\n \; 
B^\pm \rb u(\bp) \;,
\label{2.8}
\eeq
\ni
with $K=(k' \sp k)/2$.
The dynamical content of the $\p N$ interaction is carried by the functions 
$D^\pm$ and $B^\pm$ and their main properties were reviewed 
by H\"ohler \cite{H83}.
The chiral structure of these sub amplitudes was discussed
by Becher and Leutwyler \cite{BL1,BL2} a few years ago, 
in the framework of covariant perturbation theory,
and here we employ their results. 
As far as power counting is concerned, in Appendix \ref{secA:kinematics} one finds 
$[\ub(\bp')\;u(\bp)]^{(3)}\rar \cO(q^0) $ 
and $ [ \frac{i}{2m} \ub(\bp')\;\s_{\m \n}(p' \sm p)^\m K^\n \; 
u(\bp) ]^{(3)} \rar \cO(q^2)$, 
indicating that one needs the expansions of  $D^\pm$ and $B^\pm$ 
up to $\cO(q^3)$ and $\cO(q)$ respectively.

At low and intermediate energies, the $\pi$N amplitude is given by a 
nucleon pole superimposed to a smooth background. 
One then distinguishes the pseudovector (PV) Born term from 
a remainder (R) and writes
\beq
T^\pm = T_{pv}^\pm + T_R^\pm \;.
\label{2.9}
\eeq
The former contribution depends on just two observables, namely the 
nucleon mass $m$ and the $\p$N coupling constant $g$, 
as prescribed by the Ward-Takahashi identity \cite{WTI}. 
The calculation of these quantities in chiral perturbation theory may 
involve loops and other coupling constants but,
at the end, results must be organized so as to reproduce the physical 
values of both $m$ and $g$ in $ T_{pv}^\pm$ \cite{MK}.
For this reason, one uses the constant $g$, instead of  $(g_A/f_\p)$, 
since the former is indeed the observable determined by the residue 
of the nucleon pole \cite{H83,GSS,BL2}.
The $pv$ Born sub amplitudes are given by 
\bea
D_{pv}^+ &=& \frac{g^2}{2m}\; \lp \frac{k'\cd k}{s-m^2}+
\frac{k'\cd k}{u-m^2}\rp \;,
\label{2.10}\\
B_{pv}^+ &=&  - g^2 \; \lp \frac{1}{s-m^2}-\;\frac{1}{u-m^2}\rp \;,
\label{2.11}\\
D_{pv}^- &=&  \frac{g^2}{2m} \; \lp \frac{k\cd k'}{s-m^2}-\;
\frac{k\cd k'}{u-m^2}-\; \frac{\n}{m} \rp \;,
\label{2.12}\\
B_{pv}^- &=& - g^2 \; \lp \frac{1}{s-m^2}+ \frac{1}{u-m^2} +
\frac{1}{2m^2} \rp \;,
\label{2.13}
\eea
\ni
where $s$ and $u$ are the usual $\p N$ Mandelstam variables.
In the case of free pions, their chiral orders are respectively 
$[D_{pv}^+, B_{pv}^+, D_{pv}^- , B_{pv}^-] \rar \cO[q^2,\, q^{-1}\,, q,\, q^0]$,
but important changes do occur when the pions become off-shell.

The amplitudes $T_R^\pm$ receive contributions from both tree interactions 
and loops.
The former can be read directly from the basic lagrangians and correspond 
to polynomials in  $t=(k' \sm k)^2$ and $\n=(p' \sp p)\cd (k' \sp k)/4m$, 
with coefficients given by renormalized LECs \cite{BL2}. 
The latter are more complex and depend on Feynman integrals. 
In the description of $\p N$ amplitudes below threshold, one  approximates
both types of contributions by polynomials and writes \cite{H83,HJS}
\beq
X_R = \sum x_{mn} \n^{2m}t^n \;,
\label{2.14}
\eeq
\ni where $X_R$ stands for $D_R^+$, $B_R^+/\n$, $D_R^-/\n$ or $B_R^-$.
The subthreshold coefficients $x_{mn}$ have the status of observables, 
since they can be obtained by means of dispersion relations applied to 
scattering data.
As such, they constitute an important source of information about the 
values of the LECs to be used in effective lagrangians.

The isospin odd subthreshold coefficients include leading order terms, 
which implement the predictions made by Weinberg \cite{W66} 
and Tomozawa \cite{T66} for $\p N$ scattering lengths, given by
\beq
D_{WT}^- = \frac{\n}{2 f_\p^2} \;, \;\;\;\;\;\;\;\;
B_{WT}^- = \frac{1}{2 f_\p^2} \;.
\label{2.15}
\eeq
\ni
For free pions, one has $[D_{WT}^-, B_{WT}^-] \rar \cO[q,\, q^{0}]$, 
but these orders of magnitude also change when pions become virtual.

Quite generally, the ranges of nuclear interactions are determined by 
$t$-channel exchanges.
At $\cO(q^3)$, the TPE-3NP involves only single-pion 
exchanges among different nucleons and has the longest possible range.
Another $t$-channel structure becomes apparent at $\cO(q^4)$, 
associated with the pion cloud of the nucleon, which gives rise to both 
scalar and vector form factors \cite{GSS}.
These effects extend well beyond 1 fm \cite{R01,IG} and a 
limitation of the power series given by Eq. (\ref{2.14}) is that they cannot 
accommodate these ranges, since Fourier transforms of polynomials yield 
only $\d$-functions and its derivatives. 
In the description of the $\p N$ amplitude produced by Becher and 
Leutwyler \cite{BL2}, one learns that the only sources of medium range 
($mr$) effects are their diagrams $k$ and $l$,
which contain two pions propagating in the $t$-channel. 
In our derivation of the TPE-3NP, the loop content of these  
diagrams is not approximated by power series and, for free pions, 
the non-pole subamplitudes are written as 
\bea
D_R^+ &=& D_{mr}^+(t)  + \lb \db_{00}^+ + d_{10}^+ \n^2 + \db_{01}^+ t \rb_{(2)} 
+ \lb d_{20}^+ \n^4 + d_{11}^+ \n^2 t  + \db_{02}^+ t^2 \rb_{(3)} \;,
\label{2.16}\\[2mm]
B_R^+ &=& B_{mr}^+(t) +  \lb b_{00}^+ \n \rb_{(1)} \;,
\label{2.17}\\[2mm]
D_R^- &=&  D_{mr}^-(t) + \lb \n / (2f_\p^2) \rb_{(1)} 
+ \lb  \db_{00}^- \n + d_{10}^- \n^3 
+ \db_{01}^- \n t \rb_{(3)} \;,
\label{2.18}\\[2mm]
B_R^- &=&  B_{mr}^-(t) + \lb 1/(2 f_\p^2) + \bb_{00}^-  
\rb_{(0)}+ \lb b_{10}^- \n^2 + 
\bb_{01}^- t \rb_{(1)} \;,
\label{2.19}
\eea
\ni
where the labels $(n)$ outside the brackets indicate the presence of 
$\cO (q^n)$ leading terms and $mr$ denotes terms associated with the 
nucleon pion cloud. 
The {\em bar} symbol over some coefficients indicates that they do not 
include both Weinberg-Tomozawa and medium range contributions, 
which are accounted for explicitly.
The functions $D_R^\pm$ and $B_R^\pm$ depend on the parameters 
$f_\p$, $g_A$, $\m$, $m$ and on the LECs $c_i$ and $\bar{d}_i$,
which appear into higher order terms of the effective lagrangian. 
The subthreshold coefficients are the door through which LECs enter our 
calculation and their explicit forms are given in Appendix \ref{secA:coefficients}.

%\newpage
%fig.3^^^^^^^^^^^^^^^^^^^^^^^^^^^^^^^^^^^^^^^^^^^^^^^^^^^^^^^^^^^^^^^^^^^^^^^^^
\begin{figure}[ht]
\begin{center}
%\vspace{-5mm}
%\hspace*{-25mm}
\includegraphics[width=1.0\columnwidth,angle=0]{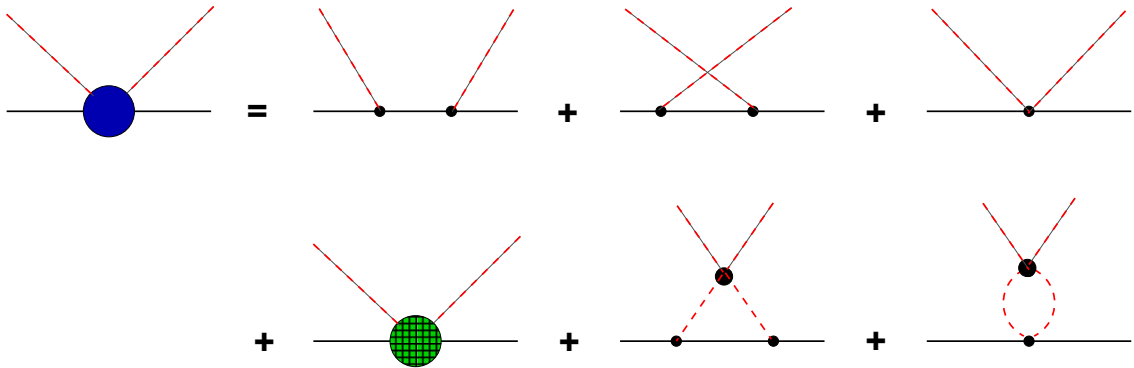}
%\vspace{-15mm}
\caption{(Color online) Representation of the $\p N$ amplitude used 
in the construction of the TPE-3NP.} 
\label{F3}
\end{center}      
\end{figure}
%^^^^^^^^^^^^^^^^^^^^^^^^^^^^^^^^^^^^^^^^^^^^^^^^^^^^^^^^^^^^^^^^^^^^^^^^^^^^^^

%\vspace{-5mm}

The dynamical content of the $\cO(q^3)$ $\p N$ amplitude is shown in Fig. \ref{F3}.
The first two diagrams correspond to $PV$ Born amplitudes,
whereas the third one represents the Weinberg-Tomozawa contact interaction,
all of them with physical masses and coupling constants. 
The fourth graph summarizes the terms within square brackets in 
Eqs. (\ref{2.16}-\ref{2.19}) and depends on the LECs.
Finally, the last two diagrams describe medium range effects owing to the 
nucleon pion cloud, associated with scalar and vector form factors. 
This decomposition of the $\p N$ amplitude has also been used in our derivation
of the two-pion exchange components of the $NN$ interaction \cite{HR, HRR}
and hence the present calculation is consistent with those results.

%^^^^^^^^^^^^^^^^^^^^^^^^^^^^^^^^^^^^^^^^^^^^^^^^^^^^^^^^^^^^^^^^^^^^^^
%3333333333333333333333333333333333333333333333333333333333333333333333
\section{intermediate $\p N$ amplitude}
\label{sec:piN-amplitude}

\begin{figure}[ht]
\begin{center}
%\vspace{40mm}
%\hspace*{-25mm}
\includegraphics[width=1.0\columnwidth,angle=0]{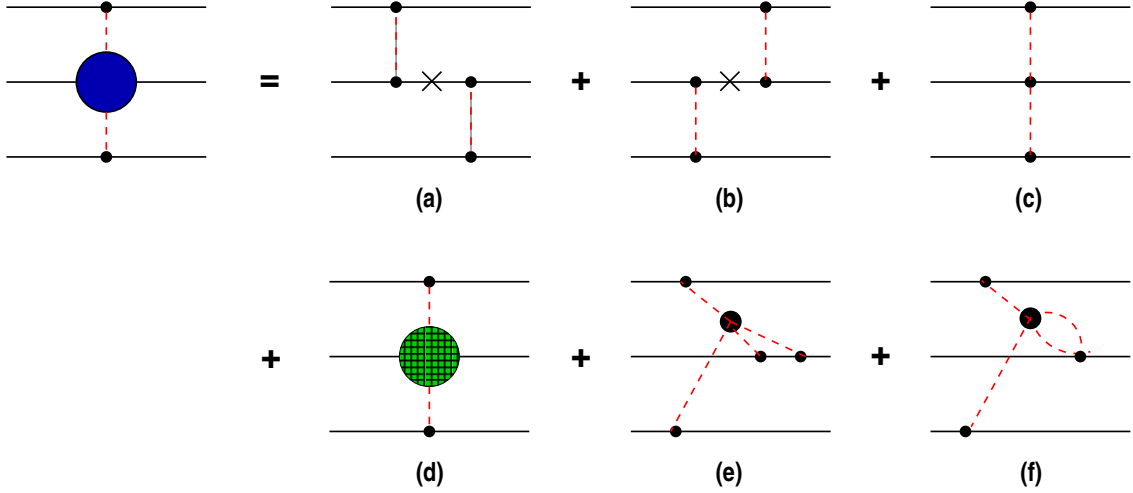}
%\vspace{-20mm}
\caption{(Color online) Structure of the $\cO(q^4)$ two-pion exchange 
three-nucleon potential} 
\label{F4}
\end{center}      
\end{figure}

The combination of Figs. \ref{F2} and \ref{F3} gives rise to the TPE-3NP, 
associated with the six diagrams shown in Fig. \ref{F4}.
In the sequence, we discuss their individual contributions to the 
subamplitudes $D^\pm$ and $B^\pm$.
We are interested only in the longest possible component of the potential and
numerators of expressions are systematically simplified by using 
$k^2 \rar \m^2$ and $k^{'2} \rar \m^2$.
In configuration space, this corresponds to keeping only those terms which 
contain two Yukawa functions and neglecting interactions associated with 
Figs. \ref{F1} (c) and \ref{F1} (d).

%\newpage
%fig.4^^^^^^^^^^^^^^^^^^^^^^^^^^^^^^^^^^^^^^^^^^^^^^^^^^^^^^^^^^^^^^^^^^^^^^^^^
%^^^^^^^^^^^^^^^^^^^^^^^^^^^^^^^^^^^^^^^^^^^^^^^^^^^^^^^^^^^^^^^^^^^^^^^^^^^^^^

\ni
{\bf $\bullet$ diagrams $(a)$ and $(b)$: }
The crosses in the nucleon propagators of Figs. \ref{F4} (a) and \ref{F4} (b) indicate that they 
do not include forward propagating components, so as to avoid double 
counting when the potential is used in the dynamical equation.
The covariant evaluation of these contributions is based on 
Eqs. (\ref{2.10}-\ref{2.13}). 
Denoting by $\pb$ the momenta of the propagating nucleons, 
the factors $1/(s \sm m^2)$ and $1/(u \sm m^2)$ are decomposed as
\beq
\frac{1}{ (\pb^0)^2 \sm \bar{E}^2} = \frac{1}{2 \bar{E}(\pb^0 \sm \bar{E})}
- \frac{1}{2 \bar{E}(\pb^0 \sp \bar{E})} \;,
\label{3.1}
\eeq

\ni
with $\bar{E}=\sqrt{m^2 \sp \bar{\bp}^2}$.
The first term represents forward propagating nucleons, 
associated with the iteration of the OPEP,
whereas the second one gives rise to connected contributions.
Discarding the former and using the results of Appendix \ref{secA:kinematics}, one has
\beq
1 /(\lc_u^s\rc - m^2) \rar 
- 1 / \lb 4m^2 + \lp 3 \bq_r^2 \sp \bq_\r^2/3  \sp 16 \bQ_\r^2/3
\pm 10 \bq_r \cd \bQ_\r /\sqrt{3} \mp  2\bq_\r \cd \bQ_r /\sqrt{3} \rp \rb \;.
\label{3.2}
\eeq
After appropriate truncation, one obtains
\bea
&& D_{a b}^+ = -\,\frac{g^2}{8m^3}\, (2\m^2 \sm t)  \rar \cO(q^2) \;,
\label{3.3}\\[2mm]
&& B_{a b}^+ \rar \cO(q^2) \;,
\label{3.4}\\[2mm]
&& D_{a b}^- = - \frac{g^2}{2m^2} \,\n \rar \cO(q^2) \;,
\label{3.5}\\[2mm]
&& B_{a b}^- \rar \cO(q^2) \;,
\label{3.6}
\eea

\ni
where we have used the fact that, in the case of virtual pions, 
$\n \rar \cO(q^2)$.

\ni
{\bf $\bullet$ diagrams $(c)$ and $(d)$: }
These contributions are purely polynomial, can be read directly 
from Eqs. (\ref{2.16}-\ref{2.19}),
and are given by 
\bea
&& D_{c d}^+ = - \frac{4\, c_1}{f_\p^2}\,\m^2 
+ \lb \frac{c_3}{f_\p^2} + \frac{g_A^4\m}{16 \,\p f_\p^4} 
\rb (2\, \m^2 \sm t) \rar \cO(q^2)  \;,
\label{3.7}\\[2mm]
&& B_{c d}^+ \rar \cO(q^2) \;,
\label{3.8}\\[2mm]
&& D_{c d}^- =   \frac{1}{2 f_\p^2} \,\n \rar \cO(q^2) \;,
\label{3.9}\\[2mm]
&& B_{c d}^- = \frac{1}{2 f_\p^2} +  \frac{2\, c_4  m}{f_\p^2} 
-  \frac{g_A^4 m \m}{8 \,\p f_\p^4}\rar \cO(q^0) \;.
\label{3.10}
\eea

\ni
{\bf $\bullet$ diagrams $(e)$ and $(f)$: }
The medium range components of the intermediate $\p N$ amplitude are  
\bea
&& D_e^+ = \frac{g_A^2 \m}{64\p^2 f_\p^4}\,(2t \sm \m^2) 
\lb (1\sm t/2\m^2) \,\Pi_t - 2\p\rb \rar \cO(q^3) \;,
\label{3.11}\\[2mm]
&& D_{ef}^+ \rar \cO(q^4) \;,
\label{3.12}\\[2mm]
&& B_e^- = \frac{g_A^2 m \m}{16\p^2 f_\p^4} \lb (1\sm t/4\m^2) 
\,\Pi_t - \p\rb \rar \cO(q) \;,
\label{3.13}
\eea
\ni
where $\Pi_t$ is the dimensionless Feynman integral 
\beq
\Pi_t = \int_0^1 da\; \frac{\m^2\, F(a)}{t \sm M^2} \;\;\;
\leftarrow \;\;\; M=2\m /a \;,
\;\;\; F(a) = \frac{8}{a^2} \, \tan^{-1} \lb \frac{ma\,
\sqrt{1 \sm a^2}}{\m \, (1\sm a^2/2)}\rb \;.
\label{3.14}
\eeq
The amplitude $D_{ef}^-$, proportional to $\n$, is $\cO(q^3)$ for free pions 
and here becomes $\cO(q^4)$.
Thus, in fact, diagram $(f)$ does not contribute to the TPE-3NP at $\cO(q^4)$.

\ni
{\bf $\bullet$ full results: }
The Golberger-Treiman relation $g/m=g_A/f_\p$ is valid up to $\cO(q^2)$ and
can be used in diagrams $(a)$ and $(b)$.
One then has
\beq
D^+ = \frac{\s(2\m^2)}{f_\p^2}
+ \frac{(2\m^2 \sm t)}{f_\p^2} \lb - \frac{g_A^2}{8m} + c_3 
+ \frac{g_A^2 (1 \sp g_A^2) \m}{16 \p f_\p^2}
- \frac{g_A^2 \m}{128 \p^2 f_\p^2} \, (1 \sm 2t/\m^2) \,\Pi_t \rb \;,
\label{3.15}
\eeq
\ni
where
\beq
\s(t = 2\m^2) = -4 \, c_1\,\m^2 - \frac{3g_A^2 \m^3}{32\p f_\p^2} 
\label{3.16}
\eeq
\ni
is the value of the scalar form factor at the Cheng-Dashen point \cite{BL1}.
The remaining amplitudes read
\bea
&& B^+ \rar \cO(q^2) \;,
\label{3.17}\\[2mm]
&& D^- =   \frac{1 \sm g_A^2}{2 f_\p^2} \,\n  \;,
\label{3.18}\\[2mm]
&& B^- =   \frac{1+ 4\, c_4  m}{2 f_\p^2} 
-  \frac{g_A^2(1\sp 2 g_A^2) m \m}{16 \,\p f_\p^4} 
+ \frac{g_A^2 m \m}{16\,\p^2 f_\p^4} \,  (1 \sm t/4\m^2)\, \Pi_t  \;.
\label{3.19}
\eea
The subamplitudes $D^\pm$ and $B^\pm$ begin at $\cO(q^2)$ and 
one needs just the leading terms in the spinor matrix 
elements of Eq. (\ref{2.8}), which is rewritten as
\bea
&& T^+ = 2m\, D^+ \;,
\label{3.20}\\[2mm]
&& T^- = 2m\, D^- + \,i \, \bsig^{(3)} \cd \bk' \st \bk \, B^- \;,
\label{3.21}\eea 
\ni
with $D^+ \rar \cO(q^2) \sp \cO(q^3)$, $D^- \rar \cO(q^2)$, and 
$B^- \rar \cO(q^0) \sp \cO(q)$.

\ni
{\bf $\bullet$ $\cO(q^3)$ reduction: }
In order to compare our amplitudes with previous $\cO(q^3)$ results, 
one notes that, in case corrections are dropped, one would have
\bea
&& D^+ = \frac{\s(0)}{f_\p^2}
+ \frac{(2\m^2 \sm t)}{f_\p^2} \lc - \lb \frac{g_A^2}{8m}\rb  + c_3 \rc \;,
\label{3.22}\\[2mm]
&& B^- =   \lb \frac{1}{2 f_\p^2} \rb + \frac{2\, c_4  m}{f_\p^2} \;.
\label{3.23}
\eea
These expressions agree with those derived directly from a chiral 
lagrangian \cite{Oq3}, except for the terms within square brackets in both 
$D^+$ and $B^-$.
The former corresponds to a Born contribution whereas the latter is due
to diagram (c) in Fig. \ref{F4}, associated with the Weinberg-Tomozawa term.

%^^^^^^^^^^^^^^^^^^^^^^^^^^^^^^^^^^^^^^^^^^^^^^^^^^^^^^^^^^^^^^^^^^^^^
%444444444444444444444444444444444444444444444444444444444444444444444
\section{two-pion exchange potential}
\label{sec:TPE-P}

The expansion of the TPE-3NP up to $\cO(q^4)$ requires only leading terms 
in vertices and propagators.
In order to derive the non-relativistic potential in momentum space, 
one divides Eq. (\ref{2.7})  
by the relativistic normalization factor $\sqrt{2E}\simeq \sqrt{2m}$ 
for each external nucleon leg and 
writes\footnote{One notes that this expression is identical 
with Eq. (33) of Ref. \cite{CDR}
divided by $8m^3$.}
\bea
&& \bar{t}_3 = \frac{g_A^2}{4f_\p^2}\,
\frac{1}{\bk^2 \sp \m^2}\;\frac{1}{\bk^{'2}\sp \m^2}\;
\bsig^{(1)}\cd \bk \; \bsig^{(2)}\cd \bk' \;
\nn\\[2mm]
&&
\times \lb \btau^{(1)}\cd \btau^{(2)}\, D^+ 
- i\, \btau^{(1)}\times \btau^{(2)} \cd \btau^{(3)} 
\lp D^- + \frac{i}{2  m}\, \bsig^{(3)} \cd \bk' \st \bk \; B^- \rp \rb \;.
\label{4.1}
\eea
The configuration space potential has the form 
\beq
V_3(\br, \bro) =  \btau^{(1)}\cd \btau^{(2)} \, V_3^+(\br, \bro) 
+ \btau^{(1)}\times \btau^{(2)} \cd \btau^{(3)} \, V_3^-(\br, \bro) 
+ \mathrm{cyclic \; permutations},
\label{4.2}
\eeq
\ni
with 
\bea
&&\!\!\!\!\!\!\!\!\!\! V_3^+(\br, \bro) = C_1^+ \; \su \cd \hrtu \, 
\sd \cd  \hrdt \; U_1(\rtu) \,U_1(\rdt) 
\nn\\[2mm]
&& + \; C_2^+ \, \lc 
(1/9)\,\su \cd \sd \lb U(\rtu) \sm U_2(\rtu) \rb \, 
\lb U(\rdt) \sm U_2(\rdt) \rb \right.
\nn\\[2mm]
&& + \left.
(1/3)\, \su \cd \hrdt \, \sd \cd \hrdt \, 
\lb U(\rtu) \sm U_2(\rtu) \rb \, U_2(\rdt)  \right.
\nn\\[2mm]
&& + \left. 
(1/3) \, \su \cd \hrtu \, \sd \cd \hrtu \, U_2(\rtu) \, 
\lb U(\rdt) \sm U_2(\rdt) \rb \right.
\nn\\[2mm]
&& + \left. 
\su \cd \hrtu \, \sd \cd \hrdt \; \hrtu \cd \hrdt \; 
U_2(\rtu) \, U_2(\rdt) \rc
\nn\\[2mm]
&& + \; C_3^+ \, \su \cd \bnb_{31}^I \, \sd \cd \bnb_{23}^I \; 
\bnb_{31}^I \cd \bnb_{23}^I \, \lb I^0 - 2\, I^1 \rb \;,
\label{4.3} 
\eea
\bea
&&\!\!\!\!\!\!\!\!\!\! V_3^-(\br, \bro)  = 
C_1^- \lc 
(1/9)\, \su \st \sd \cd \str \,\lb U(\rtu) \sm U_2(\rtu) \rb \, \lb U(\rdt) 
\sm U_2(\rdt) \rb \right.
\nn\\[2mm]
&& + \left. 
(1/3)\, \str \st \su \cd \hrdt \, \sd \cd \hrdt \, \lb U(\rtu) 
\sm U_2(\rtu) \rb \, U(\rdt) \right.
\nn\\[2mm]
&& + \left. 
(1/3)\, \su \cd \hrtu \, \sd \st \str \cd \hrtu \; U_2(\rtu) \, 
\lb U(\rdt) \sm U_2(\rdt) \rb \right. 
\nn\\[2mm]
&& + \left. 
\su \cd \hrtu \; \sd \cd \hrdt \; \str \cd \hrtu \st \hrdt 
\; U_2(\rtu) \, U_2(\rdt) \rc
\nn\\[2mm]
&& \!\!\!\!\!\!\!\!\!\! + \; C_2^- \lc
\su \cd \lp i \bnb_{\! 31}^{w\!f} \sm i \bnb_{\! 23}^{w\!f} \rp \; 
\sd \cd \hrdt \, \lb U(\rtu) \sm U_2(\rtu) \rb \, U_1(\rdt)  \right.
\nn\\[2mm]
&& + \left. 
\su \cd \hrtu \,
\sd \cd \lp i \bnb_{\! 31}^{w\!f} \sm i \bnb_{\! 23}^{w\!f} \rp \, 
U_1(\rtu) \, \lb U(\rdt) \sm U_2(\rdt) \rb \right.
\nn\\[2mm]
&& + \left.
3 \, \su \cd \hrtu \, \sd \cd \hrdt \, 
\lp i \bnb_{\! 31}^{w\!f} \sm i \bnb_{\! 23}^{w\!f} \rp \cdot 
\lb \hrtu \ U_2(\rtu) \, U_1(\rdt) + \hrdt \, 
U_1(\rtu) \, U_2(\rdt) \rb \rc 
\nn\\[2mm]
&& \!\!\!\!\!\!\!\!\!\! + \;  C_3^- \; 
\su \cd \bnb_{\! 31}^I\, \sd \cd \bnb_{\! 23}^I \, 
\str \cd \bnb_{\! 31}^I \st \bnb_{\! 23}^I \, \lb I^0 - I^1 /4 \rb \;.
\label{4.4}\\[2mm]
\nn
\eea
The profile functions are written in terms of the dimensionless variables 
$\bx_{ij} = \m \, \br_{ij}$ 
and read 
\bea
&& U(x) = \frac{e^{-x}}{x} \;,
\label{4.5}\\[2mm]
&& U_1(x) = - \lp 1 + \frac{1}{x} \rp \, \frac{e^{-x}}{x} \;,
\label{4.6}\\[2mm]
&& U_2(x) = \lp 1 + \frac{3}{x} + \frac{3}{x^2} \rp  \, \frac{e^{-x}}{x} \;,
\label{4.7}\\[2mm]
&& I^n = - \; \frac{16 \p}{\m^2}\, \int \frac{d\bk}{(2\p)^3}
\, \frac{d\bk'}{(2\p)^3}\;
e^{i (\bk \cdot \br_{31} \sp \bk' \cdot \br_{23})} \lb \frac{t}{\m^2}\rb^n \, 
\frac{1}{\bk^2 \sp \m^2}\, \frac{1}{\bk'^2 \sp \m^2}\, \Pi_t(t) \;.
\label{4.8}
\eea
The last function involves the loop integral given in Eq. (\ref{3.14}) and
is discussed further in Appendix \ref{secA:I-function}.
The gradients $\bnb_{ij}^I$ act on the functions $I^n$, whereas the 
$\bnb_{ij}^{w\!f}$ act {\em only} on the wave function and give rise to 
non-local interactions, as discussed in Appendix \ref{secA:non-local}.

The strength coefficients are the following combinations of the basic 
coupling constants
\bea
&& C_1^+ = \frac{g_A^2 \m^4}{64\, \p^2 f_\p^4} \, \s(2\m^2) \;,
\label{4.9}\\[2mm]
&& C_2^+ = \frac{g_A^2 \m^6}{32\, \p^2 f_\p^4\, m} 
\lp - \frac{g_A^2}{8} + m\, c_3 
+ \frac{g_A^2(1 \sp g_A^2) m \m}{16 \p f_\p^2} \rp \;,
\label{4.10}\\[2mm]
&& C_3^+ =  \frac{g_A^4 \m^7}{4096 \, \p^3 f_\p^6}  \;,
\label{4.11}\\[2mm]
&& C_1^- = \frac{g_A^2 \m^6}{256\, \p^2 f_\p^4\, m} 
\lp 1 + 4m\, c_4 - \frac{g_A^2 (1 \sp 2 g_A^2) m \m}{8 \p f_\p^2} \rp \;,
\label{4.12}\\[2mm]
&& C_2^- = \frac{ g_A^2  (g_A^2 \sm 1) \,\m^6}{768\, \p^2 f_\p^4\,m}\;,
\label{4.13}\\[2mm]
&& C_3^-  = - \; \frac{g_A^4 \m^7}{2048 \, \p^3 f_\p^6}\;.
\label{4.14}
\eea

%^^^^^^^^^^^^^^^^^^^^^^^^^^^^^^^^^^^^^^^^^^^^^^^^^^^^^^^^^^^^^^^^^^^^
%55555555555555555555555555555555555555555555555555555555555555555555
\section{strength coefficients}
\label{sec:coefficients}

The strength constants of the potential involve a blend of four well 
determined parameters, namely $m=938.28$ MeV, $\m=139.57$ MeV, $g_A = 1.267$ 
and $f_\p = 92.4$ MeV, with  the scalar form factor at the Cheng-Dashen point 
and the LECs $c_3$ and $c_4$, which are less precise.
As far as $\s(2\m^2)$ is concerned, we rely on the results \cite{GLS}
$\s(2\m^2) - \s(0)= 15.2 \pm 0.4$ MeV, $\s(0)= 45 \pm 8$ MeV, and adopt the 
central value $\s(2\m^2)=60$ MeV.
The values quoted for the LECs in the literature vary considerably, depending 
on the empirical input employed and the chiral order one is working at.
A sample of values is given in Table \ref{TT1}.

%table 1^^^^^^^^^^^^^^^^^^^^^^^^^^^^^^^^^^^^^^^^^^^^^^^^^^^^^^^^^^^^^^^^^^^^^^^^

\begin{table}[ht]
\begin{center}
\caption{Some values of the LECs $c_3$ and $c_4$; $m$ is the nucleon mass.} 
\begin{tabular} {ccccc}
\hline\hline
Reference	& Chiral order	& $\p N$ input	& $m\,c_3$	& $m\,c_4$ \\ \hline
\cite{ButM}& 3	& amplitude at $\n=0, t=0$	& $-5.00\pm 1.43$	
& $3.62\pm 0.04$	\\ \hline 
\cite{ButM}& 3	& amplitude at $\n=0, t =2\m^2/3$	& $-5.01\pm 1.01$ 	
& $3.62\pm 0.04$	\\ \hline
\cite{FetM1}& 3	& scattering amplitude 		& $-5.69\pm 0.04$ 	
& $3.03\pm 0.16$	\\ \hline
%
% \cite{FetM2}& 3	& scattering amplitude		& $-9.44\pm 0.06$ 	
%& $2.98\pm 0.14$	\\ \hline 
% \cite{FetM2}& 3	& scattering amplitude 		& $-1.53\pm 0.03$ 	
%& $3.76\pm 0.09$	\\ \hline 
% \cite{FetM2}& 3	& scattering amplitude 		& $-11.05\pm 0.05$ 	
%& $3.05\pm 0.11$	\\ \hline
%  
\cite{BL2}	& 4	& subthreshold coefficients	& -3.4		& 2.0	\\ \hline 
\cite{BL2}	& 4	& scattering lengths		& -4.2		& 2.3	\\ \hline 
tree	& 2	& subthreshold coefficients	& -3.6		& 2.0	\\ \hline 
this work	& 3	& subthreshold coefficients	& -4.9		& 3.3	\\ 
\hline \hline
\end{tabular}
\label{TT1}
\end{center}
\end{table}

%^^^^^^^^^^^^^^^^^^^^^^^^^^^^^^^^^^^^^^^^^^^^^^^^^^^^^^^^^^^^^^^^^^^^^^^^^^^^^^^^

Our work is based on the $\cO(q^3)$ expansion of the intermediate $\p N$ 
amplitude and,
for the sake of consistency, one must use LECs extracted at the same order.
The kinematical conditions of the three-body interaction are such 
that the variable $\n$ is $\cO(q^2)$, an order of magnitude smaller than the
threshold value, $\n =\m$.
This makes information encompassed in the subthreshold coefficients better 
suited to this problem and we use results from Appendix \ref{secA:coefficients}
in order to write
\bea
&& m\, c_3 = - m\,f_\p^2 \, d_{01}^+ - \frac{g_A^4 \;m\,\m}{16 \, \p\, f_\p^2}
- \frac{77\; g_A^2 \;m\, \m}{768\, \p\, f_\p^2} \;,
\label{5.1}\\
&& m\, c_4 = \frac{f_\p^2 \, b_{00}^-}{2} - \frac{1}{4} 
+ \frac{g_A^2(1 \sp g_A^2) \, m\, \m}{16 \, \p \, f_\p^2}  \;.
\label{5.2}
\eea
Adopting the values for the subthreshold coefficients given by 
H\"ohler \cite{H83},
namely $d_{01}^+ = 1.14 \pm 0.02\, \m^{-3}$ and 
$b_{00}^- = 10.36 \pm 0.10 \, \m^{-2}$, 
one finds the figures shown in the last row of Table \ref{TT1}.
These, in turn, produce the strength coefficients 
displayed in Table \ref{TT2}.
For the sake of comparison, we also quote values employed  
in our earlier calculation \cite{CDR} and in two TM' versions \cite{TMx} 
of the Tucson-Melbourne potential \cite{TM79}.

%table2^^^^^^^^^^^^^^^^^^^^^^^^^^^^^^^^^^^^^^^^^^^^^^^^^^^^^^^^^^^^^^^^^^^^^^^^^^
\begin{table}[h]
\begin{center}
\caption{Strength coefficients in MeV.} 
\begin{tabular} {ccccccc}
\hline\hline
reference	& $C_1^+$ & $C_2^+$ & $C_3^+$ 
& $C_1^-$ & $C_2^-$ & $C_3^-$  \\ \hline
this work	  		& 0.794 	& -2.118	& 0.034	
& 0.691 & 0.014 	& -0.067	\\ \hline
Brazil \cite{CDR}	& 0.92 	& -1.99	&	  -	
& 0.67	& 	-	& -	\\ \hline
TM'(93) \cite{TMx}		& 0.60 		& -2.05		& -	  	
& 0.58	& -	& -	\\ \hline
TM'(99) \cite{TMx}		& 0.91 		& -2.26		& -	  	
& 0.61		& -	& -	\\ 
\hline\hline
\end{tabular}
\label{TT2}
\end{center}
\end{table}
%^^^^^^^^^^^^^^^^^^^^^^^^^^^^^^^^^^^^^^^^^^^^^^^^^^^^^^^^^^^^^^^^^^^^^^^^^^^^^^^^

Changes in these parameters represent theoretical progress achieved 
over more than two decades and it is worth investigating their origins in 
some detail.
With this purpose in mind, we compare present results with those 
of our previous $\cO(q^3)$ calculation \cite{CDR}.
At the chiral order one is working here, new qualitative effects begin to show up,
associated with both loops and non-local interactions. 
They are represented by terms proportional to the coefficients 
$C_3^+$, $C_2^-$ and $C_3^-$ in Eqs. (\ref{4.3}) and (\ref{4.4}).

The $\p N$ coupling is now described by $g_A^2  \m^2 /f_\p^2 = 3.66$ whereas, 
previously, the factor $g^2 \m^2 /m^2 = 3.97$ was used.
From a conceptual point of view, the latter should be preferred, 
since $g$ is indeed the proper coupling observable.
In chiral perturbation theory, the difference between both forms is ascribed 
to the parameter $\Delta_{GT} = -2 d_{18}\m^2 /g $, 
which describes the Goldberger-Treiman discrepancy \cite{BL2}.
As this is a $\cO(q^2)$ effect, both forms of the coupling become equivalent in 
the present calculation.
On the other hand, the empirical value of $g$ is subject to larger uncertainties 
and the form based on $g_A$ is more precise.
Our present choice accounts for a decrease of $8\%$ in all parameters.

The relations $C_1^+ \lrar C_s$, $C_2^+ \lrar C_p$ and $C_1^- \lrar - C'_p$ 
allow one to compare Eqs. (\ref{4.3}) and (\ref{4.4}) with Eq. (67) of Ref. \cite{CDR}.
One notes that the latter contains an unfortunate misprint in the sign of the 
term proportional to $C'_p$, as pointed out in Ref. \cite{RC86}.
In the earlier calculation, the coefficient $C_s$ was based on a 
parameter \cite{OO} $\a_\s = 1.05 \m^{-1}$, which corresponds to 
$\s(2\m^2) = 64$ MeV.
The results of Table \ref{TT2} show that the values of  
$C_2^+$ and $C_1^-$ are rather close to those of $C_p$ and $-C'_p$.
This can be understood by rewriting Eqs. (\ref{4.10}) and (\ref{4.12}) 
in terms of the subthreshold coefficient $d_{01}^+$ and 
$b_{00}^-$ as follows 
\bea
&& C_2^+ = - \,\frac{g_A^2 \m^6}{32\, \p^2 f_\p^4\, m} 
\lp m \, f_\p^2 \, d_{01}^+ + \frac{g_A^2}{8} 
+ \lb \frac{29 g_A^2 m \m}{768 \p f_\p^2} \rb \rp \;,
\label{5.3}\\[2mm]
&& C_1^- = \frac{g_A^2 \m^6}{128\, \p^2 f_\p^4\, m} 
\lp f_\p^2 \, b_{00}^- 
+ \lb \frac{g_A^2 \, m \m}{16 \p f_\p^2}\rb \rp \;.
\label{5.4}
\eea
\ni
Numerically, this amounts to $C_2^+ = -(1.845 + 0.110 + [0.163])$ MeV and 
$C_1^- = (0.624 + [0.067])$ MeV.
The second term in the former equation was overlooked in 
Ref. \cite{CDR} and should have been considered there.
The square brackets\footnote{These factors can be traced back to loop 
diagrams in Fig. \ref{F3} and are dynamically related with the term proportional 
to $C_3^\pm$, as we discuss in Appendix \ref{secA:I-function}.}
correspond to next-to-leading order contributions
and yield corrections of about $8\%$ and $11\%$ to the leading terms in 
$C_2^+$ and $C_1^-$, 
respectively.\footnote{
When comparing the new coefficients with those in the second row  
of Table \ref{TT2}, one should also take into account the $8\%$ effect 
due to the Goldberger-Treiman discrepancy.}
As the model used in Ref. \cite{CDR} was explicitly designed 
to reproduce the subthreshold coefficients quoted by H\"ohler \cite{H83}, 
it produces the very same contributions as the first terms in 
Eqs. (\ref{5.3}) and (\ref{5.4}).

%^^^^^^^^^^^^^^^^^^^^^^^^^^^^^^^^^^^^^^^^^^^^^^^^^^^^^^^^^^^^^^^^^^^^
%66666666666666666666666666666666666666666666666666666666666666666666
%%%%%%%%%%%%%%%%%%%%%%%%%%%%%%%%%%%%%%%%%%%%%%%%%%%%%
\section{Numerical Results for Three-nucleon Systems}
%%%%%%%%%%%%%%%%%%%%%%%%%%%%%%%%%%%%%%%%%%%%%%%%%%%%%
\label{sec:results}

In order to test the effects of the TPE-3NP at $\cO(q^4)$, in this section, 
we present some numerical results of Faddeev calculations for three-nucleon bound and scattering states. 
The calculations are based on a configuration space approach, in which we solve the  Faddeev integral equations \cite{Sa86,Is03,Is07},
\begin{eqnarray}
 \Phi_3 &=& \Xi_{12,3} +
\frac1{ E +i \epsilon - H_0 - V_{12} }
\nonumber \\
&& \times \left[
 V_{12} \left( \Phi_1+\Phi_2 \right)
 + W_{3} \left( \Phi_1 + \Phi_2 + \Phi_3 \right) \right],
\nonumber \\
&&  \text{(and cyclic permutations),}
\label{eq:Fad-int}
\end{eqnarray}
where $\Xi_{12,3}$, which does not appear in the bound state problem, 
is an initial state wave function for the scattering problem,  $H_0$ is a 
three-body kinetic operator in the center of mass, $V_{12}$ is a 
nucleon-nucleon (2NP) potential 
between nucleons 1 and 2, and $W_{3}$ is the 3NP displayed in Fig. \ref{F2}. 
Partial wave states of a 3N system, in which both NN and 3N forces act, 
are restricted to those with total NN angular momenta $j \le 6$ for 
bound state calculations, and $j \le 3$ for scattering state calculations.
The total 3N angular momentum ($J$) is truncated at $J=19/2$, while 3NP is
switched off for 3N states with $J>9/2$ for scattering calculations. 
These truncation procedures are confirmed to give converged results for 
the purposes of the present work.

When just local terms are retained, $\bar{t}_3$ in Eq. (\ref{4.1})
can be cast in the conventional form \cite{TM79,TM81,CDR} 
\begin{eqnarray}
\bar{t}_3 &=& -\frac{g_A^2}{4f_\pi^2} 
  \frac{F(\bk^2)}{\bk^2+\mu^2} 
  \frac{F(\bk^{\prime 2})}{\bk^{\prime 2}+\mu^2}
     (\bsig^{(1)}\cdot\bk)(\bsig^{(2)}\cdot\bk^\prime) 
\nonumber\\
&& \times 
 \Bigl[ (\btau^{(1)}\cdot \btau^{(2)})\{ a + b (\bk \cdot\bk^\prime)
% + c(\bm{k}^2+\bm{k}^{\prime 2})
\} 
\nonumber \\
&&  - (i \btau^{(1)}\times \btau^{(2)} \cdot \btau^{(3)})
 ( i \bsig^{(3)}\cdot \bk^\prime \times \bk)d~ \Bigr],
\label{eq:t3_alt}
\end{eqnarray}
where the coefficients, $a$, $b$, and $d$ are related with our potential 
strength coefficients by 
\beq
%\lp 
[C_1^+,\; C_2^+,\; C_1^- 
%\rp
] = 
\frac{1}{(4\pi)^2} \left( \frac{g_A}{2f_\pi} \right)^2
%\lp 
[-a \mu^4,\; b \mu^6,\; - d \mu^6 
%\rp
] \;.
\eeq
%
% \begin{eqnarray}
% C_1^+ &=& -\frac{1}{(4\pi)^2} \left( \frac{g_A}{2f_\pi} \right)^2 a \mu^4
% \nonumber \\
% C_2^+ &=& \frac{1}{(4\pi)^2} \left( \frac{g_A}{2f_\pi} \right)^2 b \mu^6
% \nonumber \\
% C_1^- &=& - \frac{1}{(4\pi)^2} \left( \frac{g_A}{2f_\pi} \right)^2 d \mu^6.
% \end{eqnarray}
%
The values of the coefficients, $a$, $b$, and $d$ for the TPE-3NP
at $\cO(q^4)$ are shown in Table \ref{tab:2pe-coefs}, as BR-$\cO(q^4)$.
In this table, the values for the older version of the Brazil 
TPE-3NP, BR(83) \cite{CDR}, and the potential up to $\cO(q^3)$ given by 
Eqs. (\ref{3.22}-\ref{3.23}), 
BR-$\cO(q^3)$, are shown as well.

%%%%%%%%%%%%%%%%%%%%%%%%%%%%%%%%%%%%%%%%%%%%%%%%%%%%%%%
\begin{table}[bth]
\caption{
Coefficients $a$, $b$, and $d$ of the TPE-3NP.
\label{tab:2pe-coefs}
}
\begin{tabular}{llll}
\hline\hline
~~~~~3NP & $a~\mu$\qquad & $b~\mu^{3}$\qquad &
%              \quad$c~\mu$\quad &
  $d~\mu^{3}$\qquad \\
\hline
BR-$\cO(q^4)$  & -0.981 &  -2.617 &  -0.854 \\
BR-$\cO(q^3)$  & -0.736 & -3.483 &  -1.204 \\
BR(83)    & -1.05 & -2.29 &  -0.768 \\
\hline\hline
\end{tabular}
\end{table}
%%%%%%%%%%%%%%%%%%%%%%%%%%%%%%%%%%%%%%%%%%%%%%%%%%%%%%%%%%%%%%%%%%%%%%

In Eq.\ (\ref{eq:t3_alt}), the function $F(\bk^2)$ represents a $\pi$NN 
form factor. 
We apply a dipole form factor with the cut off mass $\Lambda$, 
$\left( \frac{\Lambda^2 - \mu^2}{\Lambda^2 + \bk^2} \right)^2$, 
which modifies the profile functions $U(x)$, $U_1(x)$, and $U_2(x)$ 
in Eqs.\ (\ref{4.5}-\ref{4.7}) as
\begin{eqnarray}
U(x) &=& \frac{e^{-x}}{x} 
  -\frac{e^{- \bar{\Lambda} x}}{x}   \left( 1+  \frac{\bar{\Lambda}^2-1}{2\bar{\Lambda}} x \right) ,
\\
U_1(x) &=& -\left(1+ \frac1{x} \right) \frac{e^{-x}}{x}
%\nonumber \\
%&&
           +\bar{\Lambda}^2 \left(1+ \frac1{\bar{\Lambda}x} \right) \frac{e^{-\bar{\Lambda}x}}{\bar{\Lambda}x}
\nonumber \\
&&     +\left( \frac{\bar{\Lambda}^2-1}{2}\right) e^{-\bar{\Lambda}x},
\\
U_2(r) &=& \left( 1+\frac3x+ \frac3{x^2} \right) \frac{e^{-x}}{x}
\nonumber \\
&& -\bar{\Lambda}^3 \left( 1+\frac3{\bar{\Lambda}x}+ \frac3{(\bar{\Lambda}x)^2} \right) \frac{e^{-\bar{\Lambda}x}}{\bar{\Lambda} x}
\nonumber \\
&& - \frac{\bar{\Lambda}(\bar{\Lambda}^2-1)}{2} \left( 1+ \frac{1}{\bar{\Lambda}x}\right) e^{-\bar{\Lambda}x},
\end{eqnarray}
with $\bar{\Lambda}=\Lambda/\mu$.

\begin{figure}[ht]
\includegraphics[scale=1.0]{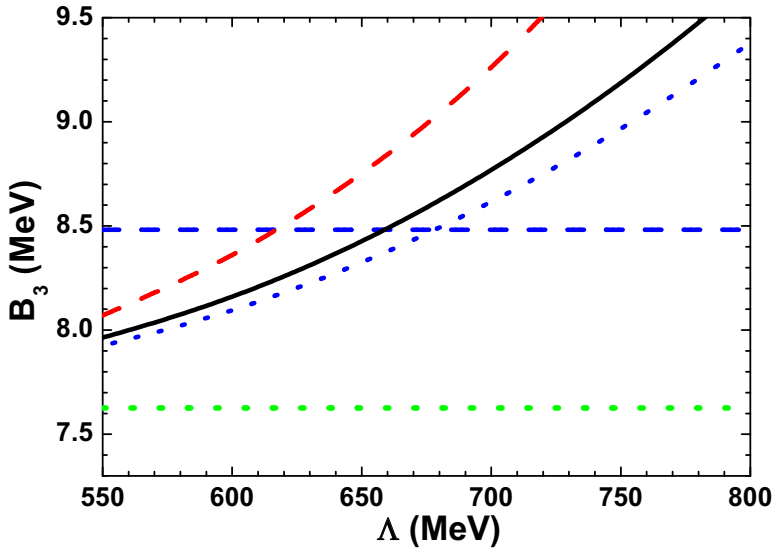}
\caption{
(Color online) The triton binding energy $B_3$ as functions of the 
cutoff mass $\Lambda$ of the $\pi NN$ dipole form factor.
The solid curve denotes the result for AV18+BR-$\cO(q^4)$,
the dashed curve for AV18+BR-$\cO(q^3)$,
and the dotted curve for AV18+BR(83).
The horizontal lines denote the AV18 result (dotted line) and the 
empirical value (dashed line).
\label{fig:be3-br-3np}
}
\end{figure}

We choose the Argonne V$_{18}$ model (AV18) \cite{Wi95} for a realistic NN 
potential, by which the triton binding energy ($B_3$) becomes 7.626 MeV, 
underbinding it by about 0.9 MeV compared to the empirical value, 8.482 MeV. 
As it is well known, the introduction of the TPE-3NP remedies this 
deficiency. 
The amount of attractive contribution depends on the cutoff mass $\Lambda$, 
as shown in Fig.\ \ref{fig:be3-br-3np}.
The solid curve shows the dependence of $B_3$ on 
$\Lambda$ for the calculation with the BR-$\cO(q^4)$ 3NP in addition to the 
AV18 2NP (AV18+BR-$\cO(q^4)$). 
In the figure, the empirical value and the AV18 result are displayed by the 
dashed and dotted horizontal lines, respectively. 
Due to the strong attractive character of the 3NP, $B_3$ is reproduced 
by choosing a rather small value of $\Lambda$, namely 660 MeV.
In the same figure, the $\Lambda$-dependence of $B_3$ for 
AV18+BR-$\cO(q^3)$ is displayed by a dashed curve 
and that for the AV18+BR(83) by a dotted curve. 
From these curves we see that AV18+BR-$\cO(q^3)$ reproduces $B_3$ for 
$\Lambda = 620$ MeV and AV18+BR(83) for $\Lambda= 680$ MeV.
In other words, the BR-$\cO(q^4)$ 3NP is slightly more attractive than the 
BR(83) 3NP and a large attractive effect occurs when one moves from the 
TPE $\cO(q^4)$ 3NP to the $\cO(q^3)$ 3NP. 
This tendency is strongly correlated with the magnitude of the coefficient 
$b$, as shown in Table \ref{tab:2pe-coefs}.
This can be understood as a dominant contribution to $B_3$ from the 
component of the TPE-3NP associated with the coefficients $b$.
This dominance is shown in Table \ref{tab:b3_cal}, 
where we tabulate calculated $B_3$ for the AV18 plus the BR-$\cO(q^4)$ 3NP and 
plus each term of the BR-$\cO(q^4)$ coming from the coefficients $a$, $b$, 
and $d$. 

%----- FIGURE  -------------------------------------------------------
%---------------------------------------------------------------------

%%%%%%%%%%%%%%%%%%%%%%%%%%%%%%%%%%%%%%%%%%%%%%%%%%%%%%%
\begin{table}[htb]
\caption{
Triton binding energy for the AV18 2NP plus the BR-$\cO(q^4)$ 3NP 
for each term of the BR-$\cO(q^4)$ 3NP with $\Lambda=660$ MeV. 
$\Delta B_3$ means the difference of the calculated binding energy 
from that of the AV18 calculation. 
\label{tab:b3_cal}
}
\begin{tabular}{lcc}
\hline\hline
 & $B_3$ (MeV)  & $\Delta B_3$ (MeV)\\
\hline
%Exp         & 8.482 \\ 
%AV18 & 7.626 \\
AV18+BR-$\cO(q^4)$  & 8.492  &  0.866 \\
AV18+BR-$\cO(q^4)$-$a$  & 7.673 & 0.047 \\
AV18+BR-$\cO(q^4)$-$b$  & 8.241 & 0.615 \\
AV18+BR-$\cO(q^4)$-$d$  & 7.787 & 0.161 \\
\hline\hline
\end{tabular}
\end{table}
%%%%%%%%%%%%%%%%%%%%%%%%%%%%%%%%%%%%%%%%%%%%%%%%%%%%%%%%%%%%%%%%%%%%%%

\begin{figure}[ht]
\includegraphics[scale=1.0]{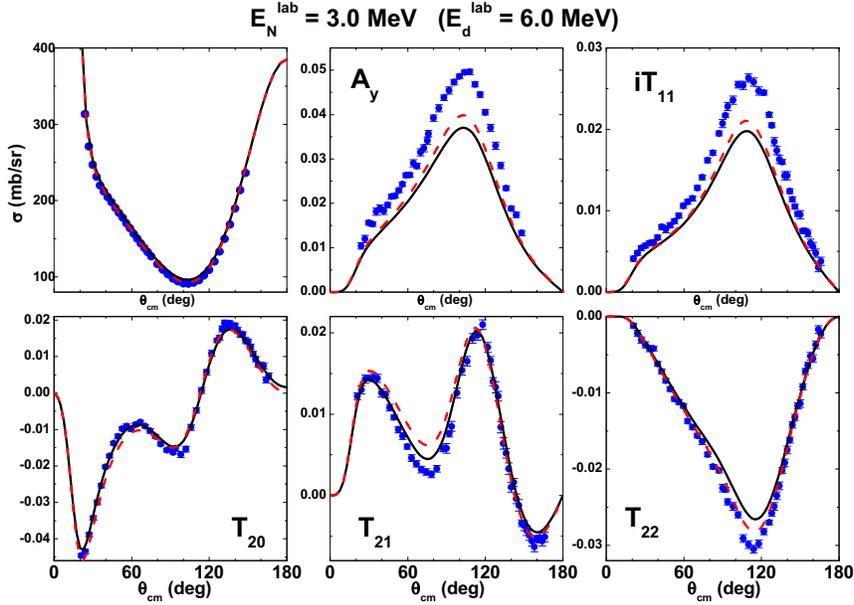}
\caption{
(Color online) 
Proton-deuteron elastic scattering observables at $E_N^{lab} =3.0$ MeV.
% differential cross sections $\sigma(\theta)$, 
%vector analyzing power of proton $A_y(\theta)$ and that of deuteron 
%$iT_{11}(\theta)$, tensor analyzing powers of 
%deuteron $T_{20}(\theta)$, $T_{21}(\theta)$, and $T_{22}(\theta)$.
Solid curves are calculations for the AV18 potential, 
and dashed curves for the AV18+BR-$\cO(q^4)$.
Experimental data are taken from Refs. \protect\cite{Sa94,Sh95}. 
\label{fig:pd3mev-br}
}
\end{figure}

In Fig. \ref{fig:pd3mev-br}, we compare six calculated observables for 
proton-deuteron elastic scattering, 
namely differential cross sections $\sigma(\theta)$, 
vector analyzing powers of the proton $A_y(\theta)$ 
and of the deuteron $iT_{11}(\theta)$, 
and tensor analyzing powers of the deuteron 
$T_{20}(\theta)$, $T_{21}(\theta)$, and $T_{22}(\theta)$, 
at incident proton energy $E_N^{lab} =3.0$ MeV, (or incident deuteron energy $E_d^{lab} = 6.0$ MeV,) 
%incident energy for proton or 6.0 MeV for deuteron, 
with experimental data of Ref.\ \cite{Sa94,Sh95}.
In the figure, the solid curves designate the AV18 calculations and the 
dashed curves the AV18+BR-$\cO(q^4)$ calculations, which are almost 
indistinguishable from the AV18+BR-$\cO(q^3)$ and AV18+BR(83) 
calculations, 
once the cut off masses are chosen so that $B_3$ is reproduced.

It is reminded that the TPE-3NF gives minor effects on the vector 
analyzing powers. 
This happens because the exchange of pions gives essentially scalar 
and tensor components of nuclear interaction in spin space, which are not 
so effective to the vector analyzing powers.
On the other hand, as is noticed in Refs. \cite{Is03b,Is04}, at $E_N^{lab} = 3.0$ MeV, 
the TPE-3NP gives a wrong contribution to the tensor analyzing power 
$T_{21}(\theta)$ around $\theta=90^\circ$.

\begin{figure}[ht]
\includegraphics[scale=1.0]{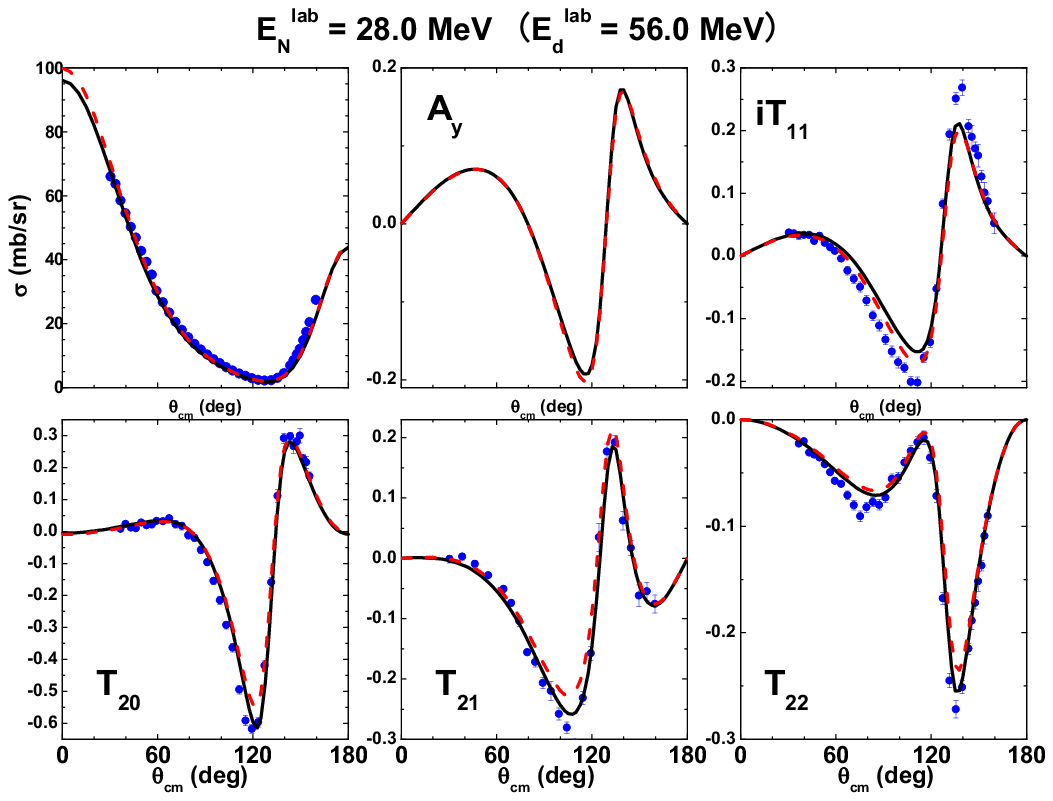}
\caption{
(Color online)
Nucleon-deuteron elastic scattering observables at $E_N^{lab} =28.0$ MeV.
Curves are calculations for neutron-deuteron scattering.
Solid curves denote calculations for the AV18 potential and dashed curves for the AV18+BR-$\cO(q^4)$.
%The meaning of the curves is same as in Fig. \protect\ref{fig:pd3mev-br}.
Experimental data are those for proton-deuteron scattering taken from Ref.\ \protect\cite{Ha84}. 
\label{fig:pd28mev-br}
}
\end{figure}

In Fig. \ref{fig:pd28mev-br}, we compare calculations of observables in 
neutron-deuteron elastic scattering at $E_N^{lab}=28.0$ MeV 
with experimental data of proton-deuteron scattering Ref. \cite{Ha84}. 
At this energy, discrepancies between the calculations and the 
experimental data in the vector analyzing power $iT_{11}(\theta)$ appear 
at $\theta\sim 100^\circ$, where $iT_{11}(\theta)$ has a minimum,  
and at $\theta \sim 140^\circ$, where  $iT_{11}(\theta)$ has a maximum, 
which are not compensated by the introduction of the TPE-3NP.
On the other hand, while the AV18 calculation almost reproduces the 
experimental data of $T_{21}(\theta)$ at $\theta \sim 90^\circ$, 
the introduction of the TPE-3NP gives a wrong effect,
as in the $E_N^{lab} = 3$ MeV case.

%----- FIGURE  -------------------------------------------------------
%---------------------------------------------------------------------

%----- FIGURE  -------------------------------------------------------
%---------------------------------------------------------------------

These results set the stage for the introduction of terms associated 
with the coefficients $C_3^+$, $C_2^-$, and $C_3^-$, 
Eqs. (\ref{4.2}-\ref{4.3}), which are new features of the 
${\cal O}(q^4)$ expansion of the TPE-3NP. 
Terms proportional to $C_3^{\pm}$, which include the rather complicated 
function $I(\bm{r}_{31},\bm{r}_{23})$ given in Appendix \ref{secA:I-function}, arise from 
a loop integral, Eq. (\ref{3.14}). 
On the other hand, the term with $C_2^-$ corresponds to a non-local potential 
and includes the gradient operator $\bm{\nabla}_{ij}^{wf}$,
which acts on the wave function and arises from the kinematical 
variable $\nu$. 
Both kinds of contributions are not expressed in the conventional local 
form shown in Eq.\ (\ref{eq:t3_alt}), which involves only the coefficients
$C_1^+$, $C_2^+$, and $C_1^-$,
and the full evaluation of their effects would require an 
extensive rebuilding of large numerical codes.
However, the coefficients of the new terms are small, and in this
exploratory paper we estimate their influence over observables as 
follows. 

The function $I(\bm{r}_{31},\bm{r}_{23})$ is approximated by Eq. (\ref{c.11}),
which amounts to replacing $\Pi_t(t)$ by a factor $-\pi$. 
Further, the kinematical factors in front of $\Pi_t(t)$ in Eqs. (\ref{3.15})
and (\ref{3.19}), namely $1-2t/\mu^2$ and $1-t/4\mu^2$, are approximately 
evaluated by putting $t\approx2 \mu^2$, which yields $-3$ and $1/2$, 
respectively.
By this procedure, the coefficients $C_3^+$ and $C_3^-$ are absorbed into 
$C_2^+$ and $C_1^-$, or in $b$ and $d$ respectively, and one has
\beq
\Delta C_2^+ = - 3 C_3^+ ,
\;\;\;\;\;\;\;\;\;\;
\Delta C_1^- = C_3^-/2.
\eeq
Numerically, this corresponds to 
$\Delta C_2^+ = -0.102~\mbox{MeV} \sim \frac1{20} C_2^+$ and 
$\Delta C_1^- = - 0.034~\mbox{MeV} \sim -\frac1{20} C_1^-$, or
$\Delta b =  -0.125 (\mu^{-3})$ and $\Delta d  = 0.042 (\mu^{-3})$.
The net change produced in the triton binding energy is
+0.026 MeV (+0.037 MeV from $\Delta C_2^+ $ and 
-0.011 MeV from $\Delta C_1^- $), just about 1/30 of 
the total increase in $B_3$ due to the local terms 
of the BR-${\cal O}(q^4)$ TPE-3NP.

The non-local term proportional to $C_2^-$ is more involved and we restrict 
ourselves to a rough assessment of its role.
We replace the variable $\nu$ by a constant $\langle \nu \rangle$ and assume,
for example, that $\langle \nu \rangle = \frac{\mu^2}{4m}$.
This changes the $C_2^-$ term in Eq. (\ref{4.4}) into the very simple form
\begin{equation}
V_3^-(\bm{r},\bm{\rho}) = C_1^- \left(\cdots\right)
 +  i \tilde{C}_2^- 
%(-i \tau^{(1)} \times \tau^{(2)} \cdot \tau^{(3)})
 \bm{\sigma}^{(1)}\cdot\hat{\bm{x}}_{31}
 \bm{\sigma}^{(2)}\cdot\hat{\bm{x}}_{23} U_1(x_{31}) U_1(x_{23})
 + C_3^- \left(\cdots \right) \;,
\end{equation}
with
\begin{equation}
\tilde{C}_2^-  =  - \frac{g_A^2}{4f_\pi^2} \frac{1-g_A^2}{2 f_\pi^2}
 \langle\nu\rangle \frac{\mu^4}{(4\pi)^2}
 =
- \frac{g_A^2(1-g_A^2) \mu^6}{512 \pi^2 f_\pi^4 m}
 = 0.021 ~\mbox{MeV} \;.
\end{equation}

Except for the isospin factor, this term is similar to that with 
$C_1^+$ (or $a$), which adds about 0.05 MeV to the triton binding energy. 
Since the potential strength $\tilde{C}_2^-$ is about 3 \% of $C_1^+$,
its contribution to the binding energy may be estimated 
to be a tiny 0.001 MeV.

%^^^^^^^^^^^^^^^^^^^^^^^^^^^^^^^^^^^^^^^^^^^^^^^^^^^^^^^^^^^^^^^^^^^^
%77777777777777777777777777777777777777777777777777777777777777777777
\section{conclusions}
\label{eq:conclusions}

In the framework of chiral perturbation theory, three-nucleon forces begin
at $\cO(q^3)$, with a long range component which is due to the exchanges 
of two pions and relatively simple.
At $\cO(q^4)$, on the other hand, a large number of different processes 
intervene and a full description becomes rather complex.
For this reason, here we concentrate on a subset of $\cO(q^4)$
interactions, namely that which still involves the exchanges of just
two pions. 
This part of the 3NP is closely related with the $\pi$N amplitude, and the 
expansion of the former up to $\cO(q^4)$ depends on the latter at $\cO(q^3)$.

Our expressions for the potential are given in Eqs. (\ref{4.2}-\ref{4.14})
and the new chiral layer of the TPE-3NP considered in this work gives rise to 
both numerical corrections to strength coefficients of already existing
terms $(C_1^+, C_2^+, C_1^-)$ and new structures in the profile functions.
Changes in numerical coefficients lay in the neighborhood of 10\% and can 
be read in Tables \ref{TT2} and \ref{tab:2pe-coefs}. 
New structures, on the other hand, arise either from loop functions 
representing form factors or the non-local terms associated with 
gradients acting on the wave function.
They correspond to the terms proportional to the parameters 
$C_3^+$, $C_2^-$ and $C_3^-$, which are small and compatible with
perturbative effects.

In order to insert our results into a broader picture, in Table \ref{TT5}
we show the orders at which the various effects begin to appear,
including the drift potential derived recently \cite{Rdrift}.

%table 5 ^^^^^^^^^^^^^^^^^^^^^^^^^^^^^^^^^^^^^^^^^^^^^^^^^^^^^^^^^^^^^^^^^^
\begin{table}[h]
\begin{center}
\caption{Chiral picture for two- and three-body forces.}
%\begin{table}[h]
\begin{tabular} {clll}
\hline\hline
beginning	  			&$\;\;\;$ TWO-BODY 		 &$\;\;\;$ TWO-BODY		 &$\;\;\;$ THREE-BODY 	\\ \hline
$\cO(q^0)$		&$\;\;\;$ OPEP: $V_T^-, V_{SS}^-$	&& 			   	  		  	 	 \\ \hline
$\cO(q^2)$		&$\;\;\;$ OPEP: $V_D^- \;\;\;\;\;$ &$\;\;\;$ TPEP: $V_C^-; V_T^+, V_{SS}^+$ &\\ \hline
$\cO(q^3)$		&&$\;\;\;$ TPEP: $V_{LS}^-, V_T^-, V_{SS}^-; V_C^+, V_{LS}^+$ 
&$\;\;\;$ TPEP: $C_1^-; C_1^+, C_2^+$ 	   		  				   			 		 \\ \hline
$\cO(q^4)$		&&$\;\;\;$ TPEP: $V_D^-; V_Q^+, V_D^+$ &$\;\;\;$ TPEP: $C_2^-; C_3^-, C_3^+$\\ 
\hline\hline
\end{tabular}
\label{TT5}
\end{center}
\end{table}
%^^^^^^^^^^^^^^^^^^^^^^^^^^^^^^^^^^^^^^^^^^^^^^^^^^^^^^^^^^^^^^^^^^^^^^^^^

The influence of the new TPE-3NP over three-body observables has been 
assessed in both static and scattering environments, adopting the 
Argonne V$_{18}$ potential for the two-body interaction. 
In order to reproduce the empirical triton binding energy, the $\cO(q^4)$
potential requires a cutoff mass of 660 MeV.
Comparing this with the value of 680 MeV for the 1983 Brazil TPE-3NP,
one learns that the later version is more attractive. 

In the study of proton-deuteron elastic scattering, 
we have calculated cross sections $\sigma(\theta)$, 
vector analyzing powers $A_y(\theta)$ of the proton  
and $iT_{11}(\theta)$ of the deuteron, 
and tensor analyzing powers 
$T_{20}(\theta)$, $T_{21}(\theta)$, and $T_{22}(\theta)$
of the deuteron, at energies of 3 and 28 MeV.
Results are displayed in Figs. \ref{fig:pd3mev-br} and \ref{fig:pd28mev-br},
where it is possible to see that there is little sensitivity to the 
changes induced in the strength parameters when one goes from $\cO(q^3)$
to $\cO(q^4)$.
Old problems, as the $A_y(\theta)$ puzzle, remain unsolved.

The present version of the TPE-3NP contains new structures, associated 
with loop integrals an non-local operators.
Their influence over observables has been estimated and found to be 
at least one order of magnitude smaller than other three-body effects.
A more detailed study of this part of the force is being carried on.

%^^^^^^^^^^^^^^^^^^^^^^^^^^^^^^^^^^^^^^^^^^^^^^^^^^^^^^^^^^^^^^^^^^^^^
\appendix
%AAAAAAAAAAAAAAAAAAAAAAAAAAAAAAAAAAAAAAAAAAAAAAAAAAAAAAAAAAAAAAAAAAAAA
\section{kinematics}
\label{secA:kinematics}

The coordinate describing the position of nucleon $i$ is $\br_i$ 
and one uses the combinations
\beq
\bR = (\br_1 \sp \br_2 \sp \br_3) /3\,, \;\;\;\;\;\;\;\;
\br = \br_2 \sm \br_1 \,, \;\;\;\;\;\;\;\;
\bro = (2\, \br_3 \sm \br_1 \sm \br_2)/\sqrt{3}\,,
\label{a.1}
\eeq
\ni
which yield
\beq
\br_1 =  \bR - \frac{\br}{2}- \frac{\bro}{2\sqrt{3}} \,, \;\;\;\;\;\;\;\;
\br_2 =  \bR + \frac{\br}{2}- \frac{\bro}{2\sqrt{3}} \,, \;\;\;\;\;\;\;\;
\br_3 =  \bR + \frac{\bro}{\sqrt{3}} \,.
\label{a.2}
\eeq
The momentum of nucleon $i$ is $\bp_i$ and one defines 
\beq
\bP = \bp_1 \sp \bp_2 \sp \bp_3 \,, \;\;\;\;\;\;\;\;
\bp_r = (\bp_2 \sm \bp_1)/2 \,, \;\;\;\;\;\;\;\;
\bp_\r = (2\, \bp_3 \sm \bp_1 \sm \bp_2)/2\sqrt{3}\,.
\label{a.3}
\eeq
Initial momenta $\bp$ and final momenta $\bp'$ are used in the combinations 
\bea
\bQ = (\bP' \sp \bP)/2 \,, &\;\;\;\;\;& \bq = (\bP' \sm \bP) \,,
\label{a.4}\\[2mm]
\bQ_r = (\bp_r' \sp \bp_r)/2 \,, &\;\;\;\;\;& \bq_r = (\bp_r' \sm \bp_r) \,,
\label{a.5}\\[2mm]
\bQ_\r = (\bp_\r' \sp \bp_\r)/2 \,, &\;\;\;\;\;& \bq_\r 
= (\bp_\r' \sm \bp_\r) \,.
\label{a.6}
\eea
In the CM, one has $\bP=0$ and the three-momenta are given by 
\bea
\bp_1 = - (\bQ_r \sm \bq_r/2) - (\bQ_\r \sm \bq_\r/2)/\sqrt{3} \;,
&\;\;\;\;& \bp'_1 = - (\bQ_r \sp \bq_r/2) - (\bQ_\r \sp \bq_\r/2)/\sqrt{3} \;,
\label{a.7}\\[2mm] 
\bp_2 = (\bQ_r \sm \bq_r/2) - (\bQ_\r \sm \bq_\r/2)/\sqrt{3} \;,
&\;\;\;\;& \bp'_2 = (\bQ_r \sp \bq_r/2) - (\bQ_\r \sp \bq_\r/2)/\sqrt{3} \;,
\label{a.8}\\[2mm] 
\bp_3 = 2 (\bQ_\r \sm \bq_\r/2)/\sqrt{3} \;,
&\;\;\;\;&  \bp'_3 =  2 (\bQ_\r \sp \bq_\r/2)/\sqrt{3} \;.
\label{a.9}
\eea
Energy conservation for on-shell particles yield the non-relativistic constraint
\beq
\bQ_r \cd \bq_r + \bQ_\r \cd \bq_\r = 0 \;.
\label{a.10}
\eeq
The momenta of the exchanged pions are written as 
\bea
k = p_1 - p'_1 \;, 
&\;\;\;\;\;&  k' = p'_2 - p_2 \;,
\label{a.11}\\[2mm]
k^0 = - (\bq_r  \sp \bq_\r /\sqrt{3}) \cd (\bQ_r \sp \bQ_\r / \sqrt{3})  /m \;,
&\;\;\;\;\;& \bk = \bq_r \sp \bq_\r /\sqrt{3} \;,
\label{a.12}\\[2mm]
k^{'0} =  (\bq_r  \sm \bq_\r /\sqrt{3}) \cd (\bQ_r \sm \bQ_\r / \sqrt{3})  /m \;,
&\;\;\;\;\;& \bk' = \bq_r \sm \bq_\r /\sqrt{3} \;,
\label{a.13}
\eea
\ni
and the Mandelstam variables for nucleon 3 read
\bea
&&\!\!\!\!\!\!\!\!\! s = (p_3 \sp k)^2 
= m^2 
- (\bq_r \sp \bq_\r/\sqrt{3}) \cdot (\bq_r \sp 2\,\bQ_r \sm \bq_\r/\sqrt{3} 
\sp 2\sqrt{3}\, \bQ_\r ) + \cO(q^4) \;,
\label{a.14}\\[2mm]
&&\!\!\!\!\!\!\!\!\! u =  (p_3 \sm k')^2 
= m^2
- (\bq_r \sm \bq_\r/\sqrt{3}) \cdot (\bq_r \sp 2\,\bQ_r \sp \bq_\r/\sqrt{3} 
\sm 2\sqrt{3}\, \bQ_\r ) + \cO(q^4) \;, 
\label{a.15}\\[2mm]
&&\!\!\!\!\!\!\!\!\! \n = (s \sm u)/4m =
- 2\, \bq_r \cd \bQ_\r /\sqrt{3} + \cO(q^4) \;.
\label{a.16}
\eea
In the evaluation of the intermediate $\p N$ amplitude, one needs
\bea
&& [\ub(\bp')\;u(\bp)]^{(3)} 
\simeq 2m + \cO(q^2) \;, 
\label{a.17}\\[2mm]
&& [ \frac{i}{2m} \ub(\bp')\;\s_{\m \n}(p' \sm p)^\m K^\n \; u(\bp) ]^{(3)} 
\simeq 2\, i\, \bsig^{(3)} \cd \bq_\r \st \bq_r  /\sqrt{3} + \cO(q^4) \;.
\label{a.18}
\eea
The $\p N$ vertex for nucleon 1 is associated with 
\beq
[\ub(\bp') \, \g_5 \, u(\bp)]^{(1)} \simeq
\bsig^{(1)} \cd (\bq_r \sp \bq_\r /\sqrt{3})  + \cO(q^3) \;,
\label{a.19}
\eeq
\ni
and results for nucleon 2 are obtained by making $\bq_r \rar - \bq_r$.

%^^^^^^^^^^^^^^^^^^^^^^^^^^^^^^^^^^^^^^^^^^^^^^^^^^^^^^^^^^^^^^^^^^^^^^^^^^^^
%BBBBBBBBBBBBBBBBBBBBBBBBBBBBBBBBBBBBBBBBBBBBBBBBBBBBBBBBBBBBBBBBBBBBBBBBBBBB
\section{subthreshold coefficients}
\label{secA:coefficients}

The polynomial parts of the amplitudes $T_R^\pm$, Eqs. (\ref{3.11}-\ref{3.16}), 
are determined by the subthreshold coefficients of Ref. \cite{BL2}.
The terms relevant to the $\cO(q^3)$ expansion are written as \cite{HR}
\bea
&& d_{00}^+ = - \frac{2\;(2c_1 - c_3)\; \m^2}{f_\p^2}+
\frac{8\; g_A^4\;\m^3}{64\; \p \;f_\p^4}
+ \lb \frac{3\;g_A^2 \; \m^3}{64\; \p \;f_\p^4}\rb_{mr}\;,
\label{b.1}\\
&& d_{01}^+ = - \frac{c_3}{f_\p^2} - \frac{48\; g_A^4 \;\m}{768\; \p\; f_\p^4}
- \lb \frac{77\; g_A^2 \; \m}{768\; \p\; f_\p^4}\rb_{mr} \;,
\label{b.2}\\
&& d_{02}^+ = \lb \frac{193\; g_A^2}{15360 \; \p \;f_\p^4\;\m}\rb_{mr} \;,
\label{b.3}\\
&& d_{00}^- = \lb \frac{1}{2\;f_\p^2}\rb_{WT} + \cO(q^2)  \;,
\label{b.4}\\
&& b_{00}^- = \lb \frac{1}{2\;f_\p^2}\rb_{WT}
+ \frac{2\; c_4 \;m}{f_\p^2}- \frac{g_A^4 \; m\; \m}{8\; \p \;f_\p^4}
- \lb \frac{g_A^2 \; m\; \m}{8\; \p \;f_\p^4}\rb_{mr} \;,
\label{b.5}\\
&& b_{01}^- =   \lb \frac{g_A^2 \; m}{96\; \p \;f_\p^4\;\m}\rb_{mr} \;,
\label{b.6}
\eea
\ni
where the parameters $c_i$ and $\tilde{d}_i$ are the usual coupling constants 
of the chiral lagrangians of order 2 and 3 respectively \cite{BKKM} and 
the {\em tilde} over the latter indicates that they were
renormalized \cite{BL2}.
Terms within square brackets labeled $(mr)$ in these results are due to 
the medium range diagrams shown in Fig. \ref{F3} and have been included 
explicitly into the functions $D_{mr}^\pm$ and $B_{mr}^\pm$. 
Terms bearing the $(W\!T)$ label were also explicitly considered
in Eqs. (\ref{2.15}-\ref{2.19}). 
The subthreshold coefficients are determined from $\p N$ scatterig data 
and a set of experimental values is given in Ref. \cite{H83}.

%^^^^^^^^^^^^^^^^^^^^^^^^^^^^^^^^^^^^^^^^^^^^^^^^^^^^^^^^^^^^^^^^^^^^^^^^^^^^^
%CCCCCCCCCCCCCCCCCCCCCCCCCCCCCCCCCCCCCCCCCCCCCCCCCCCCCCCCCCCCCCCCCCCCCCCCCCCCC
\section{functions $ I^n $}
\label{secA:I-function}

The functions $I^n$, describing loop contributions, are given by 
\beq
I^n(\br_{31}, \br_{23}) = -\; \frac{16 \p}{\m^2}\, 
\int \frac{d\bk}{(2\p)^3}\, \frac{d\bk'}{(2\p)^3}\;
e^{i (\bk \cdot \br_{31} + \bk' \cdot \br_{23})} \lb \frac{t}{\m^2}\rb^n \, 
\frac{1}{\bk^2 \sp \m^2}\, \frac{1}{\bk'^2 \sp \m^2}\, \Pi_t(t) \;.
\label{c.1}
\eeq
Using the definition Eq. (\ref{3.14}) and the Jacobi variables Eq. (\ref{a.1}), one
writes
\bea
&& I^n(\br_{31}, \br_{23}) = \lb \frac{4\,\bnb_\r^2}{3\, \m^2} \rb^n \;
I(\br_{31}, \br_{23}) \;,
\label{c.2}\\[2mm]
&& I(\br_{31}, \br_{23}) = 128 \p \, 
\int_0^1 da\; \tan^{-1} \lb \frac{ma\, \sqrt{1 \sm a^2}}{\m \, (1\sm a^2/2)}\rb \;
 L(a; \br, \bro) \;
\label{c.3}\\[2mm]
&& L(a; \br, \bro) = \int \frac{d\bq}{(2\p)^3}\, \frac{d\bQ}{(2\p)^3}\;
\frac{e^{i (\bQ \cdot \br - \sqrt{3}\,\bq \cdot \bro/2 )}} {a^2 \bq^2 \sp 4\m^2}\;
\frac{1}{[(\bQ \sm \bq)^2 \sp \m^2]}\, \frac{1}{[(\bQ \sp \bq)^2 \sp \m^2]}\,. 
\label{c.4}
\eea
The numerical evaluation of the function $L$ is can be simplified by 
using alternative representations.

\ni
{\bf $\bullet$ form 1: } One uses the Feynman procedure for manipulating 
denominators, which yields
\bea
L(a; \br, \bro) \!&=& \! 
\int_0^1 db  \int \frac{d\bq}{(2\p)^3}\, \frac{d\bQ}{(2\p)^3}\;
\frac{e^{i (\bQ \cdot \br - \sqrt{3}\,\bq \cdot \bro/2 )}}{a^2 \bq^2 \sp 4\m^2}\;
\frac{1}{[(\bQ^2 \sp \bq^2/4 \sp \m^2) \sm (1 \sm 2b)\bq \cdot \bQ ]^2 }
\nn\\[2mm]
\!&=& \! \frac{1}{8\,\p}
\int_0^1 db  \int \frac{d\bq}{(2\p)^3} \;
\frac{e^{i [(1 - 2b) \,\br - \sqrt{3}\,\bro] \cdot \bq/2}}{a^2 \bq^2 \sp 4\m^2}\;
\frac{e^{-\Theta \, r}}{\Theta} \;,
\nn\\[2mm]
\Theta \!&=& \! \sqrt{\m^2 \sp b(1\sm b)\, \bq^2} \;.
\label{c.5} 
\eea
Performing the angular integration over $\bq$, one has
\beq
L(a; \br, \bro) = 
\frac{1}{16\,\p^3}
\int_0^1 db  \int d q  \; q \,\frac{e^{-\Theta \, r}}{\Theta\,(a^2 \bq^2 \sp 4\m^2)}
\frac{\sin q \,[(1 - 2b) \,\br - \sqrt{3}\,\bro]/2}
{[(1 - 2b) \,\br - \sqrt{3}\,\bro]/2}\;.
\label{c.6} 
\eeq

\ni
{\bf $\bullet$ form 2: } The Fourier transform 
\beq
\frac{1}{\bk^2 \sp \m^2} = \int d\bx \; e^{-i \bk \cdot \bx} \; 
\frac{e^{-\m x}}{4 \p \,x}
\label{c.7}
\eeq
\ni
allows one to write  
\beq
L(a; \br, \bro) = \frac{1}{64 \p^3} \; \frac{1}{a^2} 
\int d\bz \; \frac{e^{-\m |\br_{31} \sp \bz|}}{|\br_{31} \sp \bz|}\; 
\frac{e^{-\m |\br_{23} \sm \bz|}}{|\br_{23} \sm \bz|}\; 
\frac{e^{-2 \m \, z/a}}{z }\;.
\label{c.8}
\eeq
These results may be further simplified by means of approximations.

\ni
{\bf $\bullet$ heavy baryon approximation: } In the limit $m \rar \infty$, corresponding 
to the heavy baryon case, one uses $F(a) \rar 4\p /a^2$ in Eq. (\ref{3.14}) 
and Eqs. (\ref{c.5}) and (\ref{c.7}) yield, respectively, 
\bea
&& I(\br_{31}, \br_{23}) \simeq  \frac{2}{\p} \,  
\int_0^1 \!\! db  \int_0^\infty \!\! d q  \; \lb \tan^{-1} \frac{q}{2\m} \rb 
\frac{e^{-\Theta \, r}}{\m \, \Theta}\,
\frac{\sin q \,[(1 - 2b) \,\br - \sqrt{3}\,\bro]/2}
{[(1 - 2b) \,\br - \sqrt{3}\,\bro]/2}\;,
\label{c.9}\\[2mm]
&& I(\br_{31}, \br_{23}) \simeq \frac{1}{\p} \int d\bz \; 
\frac{e^{-\m |\br_{31} \sp \bz|}}{|\br_{31} \sp \bz|}\; 
\frac{e^{-\m |\br_{23} \sm \bz|}}{|\br_{23} \sm \bz|}\; 
\frac{e^{-2 \m \, z}}{2 \m \,z^2 }\;.
\label{c.10}
\eea
\ni
{\bf $\bullet$ multipole approximation: } The integrand in Eq. (\ref{c.10}) is
peaked around $z=0$ and a multipole expansion of the Yukawa functions
produces 
\beq
I(\br_{31}, \br_{23}) \simeq  U(x_{31}) \; U(x_{23}) + \cdots .
\label{c.11}
\eeq
The same result can also be obtained by using the expansion 
$\Pi_t(t) \sim -\p [1 + t/12\m^2 + t^2/80\m^4 + \cdots]$, valid for low $t$,
directly into Eq. (\ref{c.1}).

%^^^^^^^^^^^^^^^^^^^^^^^^^^^^^^^^^^^^^^^^^^^^^^^^^^^^^^^^^^^^^^^^^^^^^^^^^^^^^
%DDDDDDDDDDDDDDDDDDDDDDDDDDDDDDDDDDDDDDDDDDDDDDDDDDDDDDDDDDDDDDDDDDDDDDDDDDDDD
\section{Non-local term}
\label{secA:non-local}

In configuration space, the variable $\bQ_\r$ corresponds to a non-local 
operator,
represented by a gradient acting on the wave function.
In order to make the dependence of $\bar{t}_3$ on $\bQ_\r$ explicit, one writes
\beq
\bar{t}_3 = [Q_\r]_i \; X_i(\bq_r, \bq_\r) \;,
\label{d.1}
\eeq
\ni
where $\bX$ is a generic three-vector, and evaluates the matrix element 
\bea
\la \psi \,|W |\psi \ra &\!=\!\!&  -\, \lb \frac{1}{(2\p)}\rb^{12}
\int d\br'\; d\bro'\; d\br\; d\bro \; \psi^*(\br', \bro') \; \psi(\br, \bro) 
\int d\bQ_r\; d\bQ_\r \; d\bq_r \; d\bq_\r \;
\nn\\[2mm]
&\!\!\times\!\!&
e^{i \lb \bQ_r \cdot (\br' \sm \br) + \; \bQ_\r \cdot (\bro' \sm \bro)  
+ \bq_r \cdot (\br' \sp \br) /2 + \bq_\r \cdot (\bro' \sp \bro) /2   \rb}
\;\; \bar{t}_3 (\bQ_r, \bQ_\r, \bq_r, \bq_\r) 
\nn\\[2mm]
&\!=\!\!& -\, \lb \frac{1}{(2\p)}\rb^{6}
\int d\br\; d\bro \; 
\lc \lb \frac{i}{2} \bnb_{\!\!\r} \, \psi^*(\br, \bro)\rb_i  \psi(\br, \bro)  
+ \psi^*(\br, \bro) \lb -\frac{i}{2} \bnb_{\!\!\r} \,\psi(\br, \bro)\rb_i \rc
\nn\\[2mm]
&\!\!\times\!\!& \int  d\bq_r \; d\bq_\r \;
e^{i \lb \bq_r \cdot \br + \bq_\r \cdot \bro \rb}
\;\; X_i (\bq_r, \bq_\r)  \;.
\label{d.2} 
\eea
This yields the potential 
\beq
V_3(\br, \bro)= -\, \frac{[2/\sqrt{3}]^3}{(2\p)^6}\,
\lb - \frac{i}{2} \, \bnblr_{\!\!\r} \,\rb_i \,
\int d\bq_r \; d\bq_\r \;
e^{i \lb \bq_r \cdot \br + \bq_\r \cdot \bro \rb}
\;\; X_i (\bq_r, \bq_\r)  \;,
\label{d.3}
\eeq
\ni
where the operator $\bnblr = \bnbr - \bnbl$ acts 
{\em  only} on the wave function.
An alternative form can be obtained by integrating Eq. (\ref{d.2}) by parts,
and one finds 
\bea
V_3(\br, \bro) &\!\!=\!\!& -\, \frac{[2/\sqrt{3}]^3}{(2\p)^6}\,
\lc \lb \int d\bq_r \; d\bq_\r \;
e^{i \lb \bq_r \cdot \br + \bq_\r \cdot \bro \rb} \;
\; \bX (\bq_r, \bq_\r) \rb \;
\cdot \lb - \,i \, \bnb_{\!\!\r}^{w\!f} \,\rb
\right.
\nn\\[2mm]
&\!\!- \!\!& \left. \lb \frac{i}{2} \, \bnb_{\!\!\r} \; \cdot 
\int d\bq_r \; d\bq_\r \;
e^{i \lb \bq_r \cdot \br + \bq_\r \cdot \bro \rb} \;\; \bX (\bq_r, \bq_\r) \rb
\rc \;.
\label{d.4}
\eea
In the case of the three-body force, the only non-local contribution
is associated with the subamplitude $D^-$, Eq. (\ref{3.18}), which yields
\bea
&& X_i = - i\, \btau^{(1)}\times \btau^{(2)} \cd \btau^{(3)} \,
\frac{1}{\bk^2 \sp \m^2}\;\frac{1}{\bk^{'2}\sp \m^2}\;
\bsig^{(1)}\cd \bk \; \bsig^{(2)}\cd \bk'
\lb \frac{g_A^2 (g_A^2-1)}{\sqrt{3}\;\;8 f_\p^4 \, m} \rb \,(\bk' \sp \bk)_i\;.
\label{d.5}
\eea
The action of $\bnb_{\!\!\r}$ on the second term of Eq. (\ref{d.4}) gives rise to an
integrand proportional to $(\bk^{'2} \sm \bk^2)$, which has short range and
does not contribute to the TPE-3NP. 
Therefore it is neglected.

%^^^^^^^^^^^^^^^^^^^^^^^^^^^^^^^^^^^^^^^^^^^^^^^^^^^^^^^^^^^^^^^^^^^^^^^^^^^
%RRRRRRRRRRRRRRRRRRRRRRRRRRRRRRRRRRRRRRRRRRRRRRRRRRRRRRRRRRRRRRRRRRRRRRRRRRR

\end{document}